\documentclass[10pt,aps,twocolumn,english,superscriptaddress,citeautoscript,showkeys,preprintnumbers,amsmath,amssymb,floatfix,footinbib,prb]{revtex4-2}
\usepackage{graphicx}
\usepackage{xcolor}
\usepackage{soul}
\usepackage{amssymb,amsmath}
\usepackage[utf8]{inputenc}
\usepackage[english]{babel}
\usepackage{txfonts}
\usepackage[font=normalsize, caption=false]{subfig}
\usepackage{hyperref}
\hypersetup{colorlinks=true, linkcolor=red, citecolor=blue}

\usepackage{xcolor}
\usepackage{listings}

\definecolor{vscode-bg}{HTML}{F2F2F2}
\definecolor{vscode-fg}{HTML}{1F1F1F}
\definecolor{vscode-comment}{HTML}{2E7D32}   
\definecolor{vscode-keyword}{HTML}{005CC5}   
\definecolor{vscode-string}{HTML}{A31515}    
\definecolor{vscode-number}{HTML}{986801}    
\definecolor{vscode-function}{HTML}{795E26}  
\definecolor{vscode-class}{HTML}{267F99}     
\definecolor{vscode-operator}{HTML}{1F1F1F}
\definecolor{vscode-line}{HTML}{6E6E6E}
\definecolor{vscode-frame}{HTML}{D0D0D0}

\lstdefinestyle{vscode-python}{
    language=Python,
    backgroundcolor=\color{gray!10},
    basicstyle=\scriptsize\ttfamily,
    keywordstyle=\bfseries\color{vscode-keyword},
    commentstyle=\itshape\color{vscode-comment},
    stringstyle=\color{vscode-string},
    numberstyle=\tiny\color{vscode-line},
    identifierstyle=\color{vscode-fg},
    showstringspaces=false,
    showspaces=false,
    keepspaces=true,
    breaklines=true,
    frame=none,
    numbers=left,
    numbersep=0.5em,
    xleftmargin=1em,
    captionpos=b,
    columns=fullflexible,    
    showtabs=false,                  
    tabsize=2,
    morekeywords={self,None,True,False,as,with,lambda},
    emph={
        set_up_clusters_and_dft, EGQCA, thermodynamic_assessment,
        average_properties, run_ml_relaxation
    },
    emphstyle=\color{vscode-function},
    numberstyle=\tiny\color{black}, 
}

\begin{document}


\title{Ab initio thermodynamic statistical modeling of the miscibility gap and the metal-insulator phase transition in SrTi\texorpdfstring{$_{1-x}$V$_x$O$_{3}$}{1-x Vx O3}}

\author{Luka Wibmer}
\email{luka.wibmer@tugraz.at}
\affiliation{Institute of theoretical and computational Physics, TU Graz, NAWI Graz, Petersgasse 16, 8010 Graz, Austria}

\author{Chiara Ostendorf}
\affiliation{Institute of theoretical and computational Physics, TU Graz, NAWI Graz, Petersgasse 16, 8010 Graz, Austria}

\author{Dominik Spath}
\affiliation{Institute of theoretical and computational Physics, TU Graz, NAWI Graz, Petersgasse 16, 8010 Graz, Austria}

\author{Christoph Heil}
\affiliation{Institute of theoretical and computational Physics, TU Graz, NAWI Graz, Petersgasse 16, 8010 Graz, Austria}

\author{Pedro N. Ferreira}
\email{nunesferreira@tugraz.at}
\affiliation{Institute of theoretical and computational Physics, TU Graz, NAWI Graz, Petersgasse 16, 8010 Graz, Austria}

\author{Markus Aichhorn}
\email{aichhorn@tugraz.at}
\affiliation{Institute of theoretical and computational Physics, TU Graz, NAWI Graz, Petersgasse 16, 8010 Graz, Austria}

\date{\today}


\begin{abstract}
The substitutional alloy SrTi$_{1-x}$V$_x$O$_3$ interpolates between the band
insulator SrTiO$_3$ and the moderately correlated metal SrVO$_3$, exhibiting a
composition-driven metal--insulator transition whose origin combines Mott
physics with local chemical disorder. 
Previous first-principles studies relied on individual supercells, which cannot
capture the thermally disordered solid solution, since configurations of
identical composition can display very different electronic properties.
Here we treat the alloy within a
generalized quasichemical approximation, a thermodynamically consistent
statistical framework in which every property is obtained as an ensemble
average over all symmetry-inequivalent clusters, weighted by occurrence
probabilities that minimize the Gibbs mixing free energy. This provides a well-defined procedure to average over different supercells, 
and places the structural and electronic
descriptions on an equal footing. From the mixing thermodynamics we obtain a
miscibility gap with a critical temperature of 1443\,K, consistent with the
limited experimental evidence for high-temperature solubility and subsequent
decomposition, and substantially larger than earlier cluster-expansion
estimates. Combining the cluster ensemble with dynamical mean-field theory, we
track the density of states at the Fermi level across the full composition
range: whereas density-functional theory alone predicts a metal for all $x>0$,
the correlated spectral function reproduces the transition, evolving from
insulating below $x\approx0.3$ to metallic near $x=1$. Finally, classifying
clusters as metallic or insulating and performing site percolation on a simple
cubic lattice yields a sharp onset of system-spanning conduction near
$x\approx0.4$.
Together, these results provide a unified, ab initio picture of the metal--insulator transition in SrTi$_{1-x}$V$_x$O$_3$
establishing a thermodynamically consistent, configuration--averaged framework applicable to the broader class of correlated materials in which substitutional disorder governs -- or drives -- electronic transitions.
\end{abstract}


\maketitle

\section{Introduction}

The interplay of strong electron--electron correlations and the
metal--insulator transition (MIT) is one of the central themes of
condensed-matter physics, underlying phenomena that range from Mott
insulators and unconventional superconductivity to colossal
magnetoresistance~\cite{imada1998, georges1996}. 
Transition-metal perovskite oxides ABO$_3$ are a particularly rich playground
for these effects~\cite{tokura2000} and are central to the emerging field of oxide electronics~\cite{ha2011,coll2019}.

SrVO$_3$ and SrTiO$_3$ are archetypal members of this family: although both adopt the same cubic perovskite structure at room temperature, SrVO$_3$ is a prototypical moderately correlated metal with a single electron in the V-$3d$ $t_{2g}$ manifold~\cite{fujimori1992,inoue1995}, whereas SrTiO$_3$, with its Ti $3d^0$ configuration, is a well-known band insulator. Substituting V for Ti in the solid solution SrTi$_{1-x}$V$_x$O$_3$ thus offers a clean route to tune the electronic structure continuously between these two limits. In experiments the system displays a composition-driven MIT, with reported critical concentrations scattered over the range $x_\mathrm{c}\approx0.4$--$0.7$, the spread partly reflecting differences in synthesis route, film thickness, and substrate~\cite{tsuiki1983,hong2002,gu2013,kanda2021,itaki2005}. This tunability makes the compound an attractive candidate for correlated transparent conductors and Mott-based devices.

The microscopic mechanism driving this transition has remained under
debate, with both Mott physics and disorder-induced (Anderson) localization
proposed as the dominant ingredient~\cite{tsuiki1983,hong2002,gu2013,kanda2021}.
Supercell density functional theory plus dynamical mean field theory (DFT+DMFT) calculations by James \emph{et al.}~\cite{James_2024} showed that the MIT depends sensitively on both local electron correlations within the $t_{2g}$ orbitals and the specific arrangement of V and Ti on the cation sublattice. More recently, site-selective x-ray spectroscopies combined with DFT+DMFT established that the Mott and Anderson characters are both essential and must be treated on an equal footing to describe the transition~\cite{laverock2025}. 

While these supercell studies capture the relevant local physics, they require
committing to a single configuration: at a given composition many
symmetry-inequivalent cells exist, frequently with markedly different electronic
properties. Even selecting the lowest-energy configuration is unsatisfactory, as
it describes an ordered ground state rather than the disordered solid solution,
which at finite temperature is stabilized by configurational entropy and is thus
realized as an ensemble of many thermally populated configurations.

Special quasi-random structures (SQS) provide a natural route to incorporate both local chemical environments and, when sufficiently large, the statistical character of substitutional disorder~\cite{liu2018}. However, the combination of large SQS cells with DFT+DMFT is computationally prohibitive, effectively restricting such approaches to less demanding treatments such as DFT+$U$~\cite{anisimov1991,dudarev1998}.

Here we overcome this limitation by describing SrTi$_{1-x}$V$_x$O$_3$ within the
generalized quasichemical approximation
(GQCA)~\cite{sher1987GQCA,ferreira2024EGQCA,EGQCA_code}, which provides a
thermodynamically consistent treatment of the disordered alloy. Each property
is obtained as an ensemble average over all symmetry-inequivalent clusters,
weighted by occurrence probabilities that follow from minimizing the Gibbs
mixing free energy, thereby replacing the need to commit to a single representative cell.

Within this unified framework, we first compute the mixing thermodynamics and the resulting miscibility gap, in agreement with the limited experimental evidence for phase separation~\cite{patino2018} and improving on earlier cluster-expansion Monte Carlo estimates~\cite{liu2018}. We then use DFT+DMFT to determine the density of states at the Fermi level across the full composition range, following the evolution of the spectral weight at $\omega=0$ as a direct measure of the electronic transition. Finally, we analyze this transition from a percolation  perspective~\cite{stauffer1994,deng2005}, linking the loss of metallic spectral weight to the connectivity of the V sublattice. Taken together, these results provide a unified, fully \emph{ab initio} picture of the coupled structural and electronic transition in SrTi$_{1-x}$V$_x$O$_3$.

\section{Model and Methods}
\textit{Generalized quasichemical approximation.}
In GQCA\cite{sher1987GQCA}, an alloy is represented as an ensemble of clusters that are both statistically and energetically independent of their atomic surroundings. Material properties at a certain composition $x$ are calculated by averaging the property over all clusters, with weights calculated by minimizing the mixing Gibbs free energy. The following section is a description of GQCA. For more details, see the formulation in Refs.~\cite{sher1987GQCA, teles2000GQCA, guilhon2015GQCA, guilhon2017GQCA, ferreira2024EGQCA}.

The material of interest is a pseudo-binary alloy A$_{1-x}$B$_x$C, where atoms of species A and B substitute one another in one or more crystallographic sites. This alloy is divided into an ensemble of clusters of a chosen size by arranging all possible clusters into $J$ nonequivalent classes with a distinct total energy $E_j$ and degeneracy $g_j$, indexed by $j = 1,2,..,J$. Depending on composition and temperature, each cluster occurs in the alloy with a certain probability $p_j$, called the cluster occurrence probability. It is defined as
\begin{equation}
    p_j = \frac{M_j}{M},
\end{equation}
with $M_j$ the number of same-class clusters with energy $E_j$ and $M = \sum^J_{j=1}M_j$ the total number of clusters.

The cluster probabilities are obtained by minimizing the Gibbs mixing free energy $\Delta G$, which is given by
\begin{equation}
    \Delta G = \Delta H - T\Delta S,
\end{equation}
with temperature $T$, the mixing enthalpy $\Delta H$ and the mixing entropy $\Delta S$.

The mixing enthalpy $\Delta H$ is expressed as
\begin{equation}
    \Delta H(x,T) = M \sum^J_{j=1} p_j\Delta_j 
\end{equation}
Here, one defines the reduced excess enthalpies $\Delta_j$ of cluster $j$ as
\begin{equation}
    \Delta_j = H_j - \frac{n-n_j}{n}H_\mathrm{A} - \frac{n_j}{n}H_\mathrm{B},
\end{equation}
where $n$ is the total number of permutable sites and $n_j$ is the number of atoms of type B in the cluster $j$.
        
The mixing entropy can be calculated from the Boltzmann definition within the microcanonical ensemble:
\begin{equation}
    \Delta S(x,T) = k_B \ln{W}.
\end{equation}
$W$ is the total number of possible alloy configurations formed with the set of clusters $M_1,M_2,..,M_J$, which is given by 
\begin{equation}
    W = \frac{N!}{N_\mathrm{A}!N_\mathrm{B}!}\frac{M!}{\prod_{J=1}^J M_j!} \prod_{j=1}^J \left(p^0_j\right)^{M_j},
    \label{W}
\end{equation}
with $N_\mathrm{A}$ and $N_\mathrm{B}$ the total number of atoms of type A and B. $p^0_j$ is the fraction of clusters of type $j$ out of all possible configurations of $N_\mathrm{A}$ and $N_\mathrm{B}$ atoms:
\begin{equation}
    p^0_j = g_j x^{n_j}(1-x)^{n-n_j},
\end{equation}
$W$ in Eq.~\ref{W} is calculated by taking the total number of ways of placing $N_\mathrm{A}$ type-A and $N_\mathrm{B}$ type-B atoms on $N$ sites $N!/(N_\mathrm{A}!N_\mathrm{B}!)$ and multiplying by the multinomial probability that this random configuration will generate the clusters $M_1,M_2,..,M_J$. This probability assumes each cluster to be independent.
Note that while in purely binary alloys $N$ is the total number of atoms, in quasi--binary alloys like SrTi$_{1-x}$V$_x$O$_3$ we need to set $N$ to only the permutable Ti/V sites.

Since both the total number of atoms $N$ and the number of clusters $M$ in a macroscopic material are very large, Stirling's approximation $\ln \alpha! = \alpha\ln\alpha - \alpha + \mathcal{O}(\ln\alpha)$ can be used to estimate $W$. With this, the Gibbs mixing free energy can be written as
\begin{equation}
    \begin{aligned}
        \Delta G(x, T) = &M \sum^J_{j=1} p_j\Delta_j \\
                        &+ N k_{B} T\left(x\ln x + (1-x)\ln(1-x)\right) \\
                        &+ M k_{B} T \sum^J_{j=1}p_j\ln\left(\frac{p_j}{p^0_j}\right).
    \end{aligned}
\end{equation}

To find the cluster probabilities $p_j(x, T)$ at temperature $T$ and composition $x$, the Gibbs mixing free energy $\Delta G$ has to be minimized with respect to $p_j$. This minimization is done with Lagrange multipliers with respect to the constraints $\sum_{j=1}^J p_j=1$ and $\sum_{j=1}^J n_jp_j=nx$. The cluster probabilities are thus given by
\begin{equation}
    p_j = \frac{g_j \eta^{n_j}e^{-\Delta_j\beta}}{\sum^J_{j=1}g_j\eta^{n_j}e^{-\Delta_j\beta}} \hspace{0.4cm} \text{with} \hspace{0.4cm} \eta = \frac{xe^{\lambda_\mathrm{L}\beta}}{1-x}, 
\end{equation}
where $\lambda_\mathrm{L}$ are the Lagrange multipliers and $\beta = (k_{B} T)^{-1}$. Minimizing the Gibbs mixing free energy results in an $n$-order polynomial equation with one unique real solution:
\begin{equation}
    \sum^J_{j=1}\left(nx-n_j\right) g_j e^{-\Delta_j\beta} \eta^{n_j} = 0.
\end{equation}

Any composition-dependent material property $\mathcal{P}$ is then calculated as a weighted average of the property $\mathcal{P}_j$ of each cluster, 
\begin{equation}
    \mathcal{P}(x,T) = \sum^J_{j=1} p_j(x,T)\mathcal{P}_j
    \label{eq:average}
\end{equation}
which can be calculated independently from each other using any ab initio method. A standard deviation of this GQCA average can by calculated as follows:
\begin{equation}
    \Delta\mathcal{P}(x,T) = \sqrt{\sum^J_{j=1} p_j(x,T)\mathcal{P}_j^2 - \left(\mathcal{P}_j(x,T)\right)^2}
\label{eq:gqca_std}
\end{equation}

\textit{Cluster generation.}
The first step in our workflow is to divide the SrTi$_{1-x}$V$_x$O$_3$ alloy into statistically and energetically independent supercells, or clusters. We use 2$\times$2$\times$2 supercells, comprising 8 formula units. This supercell size is a good compromise between computational complexity, which quickly increases for larger cells, and the ability to statistically model disorder. To generate all possible, non-equivalent clusters, the \textsc{Supercell} code~\cite{supercell} is used. In our case, there are a total of 22 non-equivalent clusters. This is a sufficient number of clusters for reliable GQCA averages~\cite{ferreira2024EGQCA}.
        
\textit{DFT calculation.}
Each of the 22 clusters is then relaxed with DFT. This is done with the \textit{Vienna Ab initio Simulation Package} (VASP), using the projector-augmented-wave (PAW) potentials \cite{vasp_paw, PAW_method}. Specifically, we use the potentials O, Sr\_sv, Ti\_sv and V from the PAW\_PBE dataset. The ionic relaxation is done such that the norm of all forces is smaller than $10^{-4}$\,eV/\AA\,and the energy difference for the electronic SCF-loop is smaller than 10$^{-7}$\,eV. The relaxed DFT energy of each cluster is used later on for the GQCA cluster probabilities.
        
After relaxation, another VASP calculation is conducted where the KS orbitals are projected to V-$d$ orbitals as a post-processing step. This is needed to perform the DMFT calculations in the next steps.
        
\textit{DMFT calculation.}
After DFT calculation, a single-shot DMFT calculation is performed to treat the highly correlated $d$ $t_{2g}$ orbitals of the V atoms more accurately. The DMFT calculation is implemented using the \textit{Toolbox for Research on Interacting Quantum Systems} (TRIQS)~\cite{TRIQS} and more, specifically, the TRIQS application DFTTools~\cite{DFTTools}, which is built for DFT+DMFT calculations.
        
To begin the DMFT setup, the non-normalized projectors generated by VASP have to be normalized using the interface tool PLOVasp to obtain the projected localized orbitals needed for the DMFT calculation. In our case, only the $t_{2g}$ orbitals are considered, since the $e_g$ orbitals are unoccupied in V. The Ti orbitals are not used in the impurity problem, since Ti, with its 3$d^0$ configuration, has no contributing $d$ states. The energy window in which the projection is done is determined for each cluster separately, using the density of states (DOS) obtained from the DFT calculation. This is done by setting the upper bound of the energy window just above the broad energy peak around the Fermi energy, which corresponds to the three bands crossing the Fermi energy with predominantly V-$t_{2g}$ character (see Fig. \ref{fig:dos_vasp_selected}). 
        
The parameters $U$ and $J$ in the Kanamori–Hubbard interaction Hamiltonian are set to $U$ = 4\,eV and $J$ = 0.65\,eV, in agreement with previous calculations on SrVO$_3$~\cite{Lechermann2006,aichhorn1010,James_2024,vaugier2012}. The double-counting correction is performed with a variant of the fully-localized limit (FLL) correction, which is used for Kanamori-type Hamiltonians~\cite{DC}.
        
The impurity model is solved with the TRIQS-based hybridization-expansion segment picture solver (CTSEG) described in Ref.~\cite{CTSEG}. The inverse temperature was set to $\beta$ = 40\,eV$^{-1}$, which is equivalent to $\approx 290$\,K.
        
After convergence, the resulting Green's function $G$ and self-energy $\Sigma$ can be used to extract important quantities for the characterization of the MIT. As in previous works~\cite{James_2024}, the spectral function at the Fermi level $A(\omega=0)$ is used for this purpose. It is calculated directly from the imaginary time Green's function $G(\tau)$ and the inverse temperature $\beta$:
\begin{equation}
    A(\omega=0) = -\frac{\beta}{\pi} G(\tau=\beta/2).
    \label{eq:A_0}
\end{equation}
$A(\omega=0)$ is evaluated on each site of a cluster independently and subsequently averaged to obtain the total cluster $A(\omega=0)$. Since the Ti sites were not considered in the DMFT calculation due to the 3d$^0$ configuration, the $A(\omega=0)$ of the Ti sites is assumed to be zero when taking the cluster average.
        
Additionally, to show the opening of a band gap around the Fermi energy during the MIT, we calculate the real-frequency spectral function $A(\omega)$ of each cluster by analytically continuing the DMFT self-energy with the maximum entropy technique, using the \textsc{LineFitAnalyzer} of the TRIQS/Maxent application~\cite{maxent}. More specifically, the auxiliary Green's function $G_{aux}(z)=(z+C-\Sigma(z))^{-1}$ with the DMFT double-counting correction taken for the constant $C$ was continued with the maximum-entropy technique. The resulting $A_{aux}(\omega)$ is then used to reconstruct the real frequency self-energy $\Sigma(\omega)$, which can then be used in the Dyson equation to calculate the local spectral function.

\section{Results}
 \textit{Structural properties.}
SrTiO$_3$ and SrVO$_3$ are isostructural and crystallize in the cubic perovskite structure belonging to the space group $Pm\overline{3}m$ (No.~221). The reported experimental lattice parameters are 3.90\,\AA~for SrTiO$_3$~\cite{yamanaka2002} and 3.84\,\AA~for SrVO$_3$~\cite{lan2003}. Our relaxed DFT lattice parameters reproduce these values very well, with relative deviations below 1\,\% for both compounds.

For the off-stoichiometric clusters, both the atomic positions and lattice parameters were fully relaxed without imposing any symmetry constraints. For some clusters, this leads to a small deviation from the ideal cubic unit cell. However, these deviations are on the order of only 0.001\,\AA, indicating that the different disordered configurations possess a strong tendency toward the cubic perovskite symmetry.

\begin{figure}[t]
    \centering
    \includegraphics[width=\linewidth]{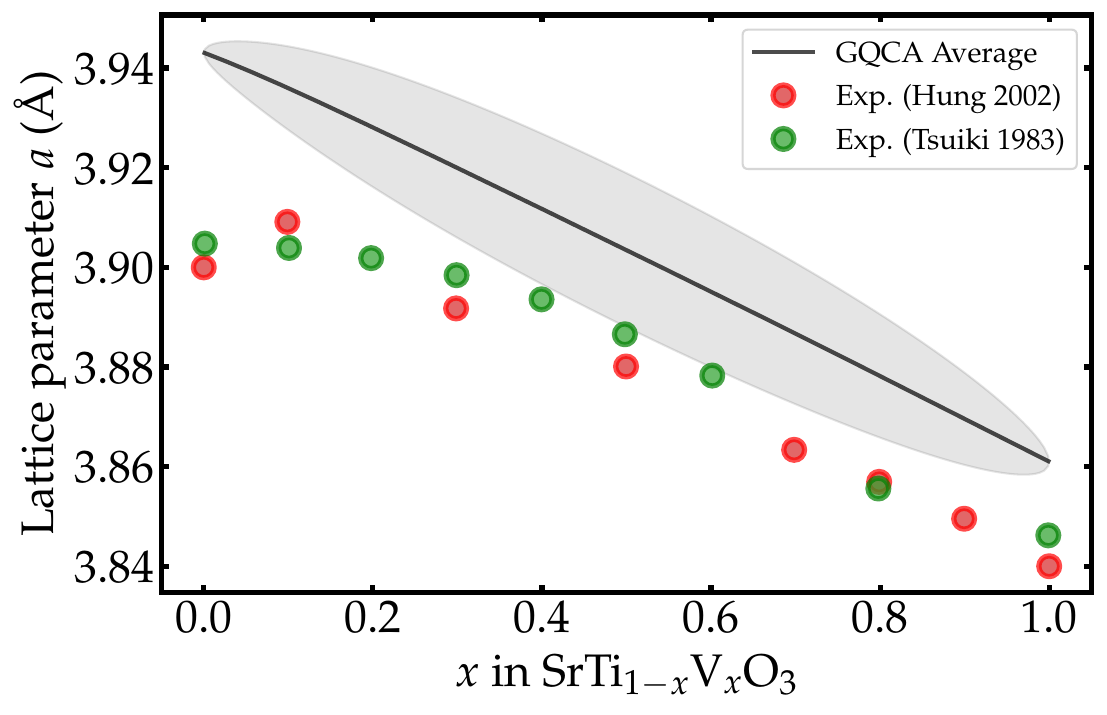}
    \caption{Lattice parameter $a$ of SrTi$_{1-x}$V$_x$O$_3$ as a function of composition. Both the GQCA average and experimental measurements \cite{hong2002, gu2013, tsuiki1983} are depicted. The gray shaded area shows the GQCA standard deviation.}
    \label{fig:lattice_parameters}
\end{figure}

The configurationally averaged lattice parameter $a$ for the solid solution SrTi$_{1-x}$V$_x$O$_3$ is obtained based on Eq.~\ref{eq:average} as
\begin{align}
a = \sum_{j=1}^{J} p_j(x, T) a_j,
\end{align}
where $a_j$ is the equilibrium lattice parameter of cluster $j$. The resulting lattice parameters are shown in Fig.~\ref{fig:lattice_parameters}. As a temperature for this and following GQCA averages we used $T=1800$\,K, which we considered to be consistent with the solid-state synthesis conditions used to prepare the solid solution from SrCO$_3$, TiO$_2$, and V$_2$O$_5$ in Ref.~\cite{hong2002}. At this constant temperature, the lattice parameter varies approximately linearly with composition, following Vegard's law~\cite{denton1991}, with no significant bowing effect. This behavior indicates that the formation of a single-phase, ideal solid solution is favorable.

Our fully \textit{ab initio} treatment closely reproduces the available experimental measurements, which report an almost linear decrease of the lattice parameter with increasing V content~\cite{hong2002, tsuiki1983} (Fig.~\ref{fig:lattice_parameters}). This behavior is consistent with the smaller ionic radius of V$^{4+}$ compared with Ti$^{4+}$. In the experiments, however, a small bowing is observed near the dilute, Ti-rich side, which is not captured by our calculations. Several factors may contribute to this deviation. First, electronic correlations were not included self-consistently during the structural relaxation. This approximation may be particularly relevant on the Ti-rich side, where dilute V impurities are embedded in a SrTiO$_3$ matrix and local V--O relaxations can be sensitive to the degree of localization of the V 3$d$ electron~\cite{liu2018}. Second, real samples may deviate from the ideal SrTi$_{1-x}$V$_x$O$_3$ solid solution. Depending on the oxygen pressure and annealing conditions, mixed V valence states, such as V$^{3+}$ or V$^{5+}$, as well as oxygen vacancies, may be present~\cite{macias2019}. These defects modify the effective B-site ionic radius and can shift the lattice parameter away from the ideal Vegard-like behavior. Finally, GGA functionals are known to overestimate bond lengths and lattice parameters in many oxides~\cite{perdew2008,evarestov2011,garcia2012}. More advanced exchange-correlation functionals, such as meta-GGAs~\cite{sun2015,furness2020} or hybrid functionals~\cite{krukau2006}, may reduce this error~\cite{evarestov2011}, although their computational cost would make their use very demanding, and in practice prohibitive within the present DFT+DMFT workflow.

\begin{figure*}[t]
    \centering
    \includegraphics[width=0.95\linewidth]{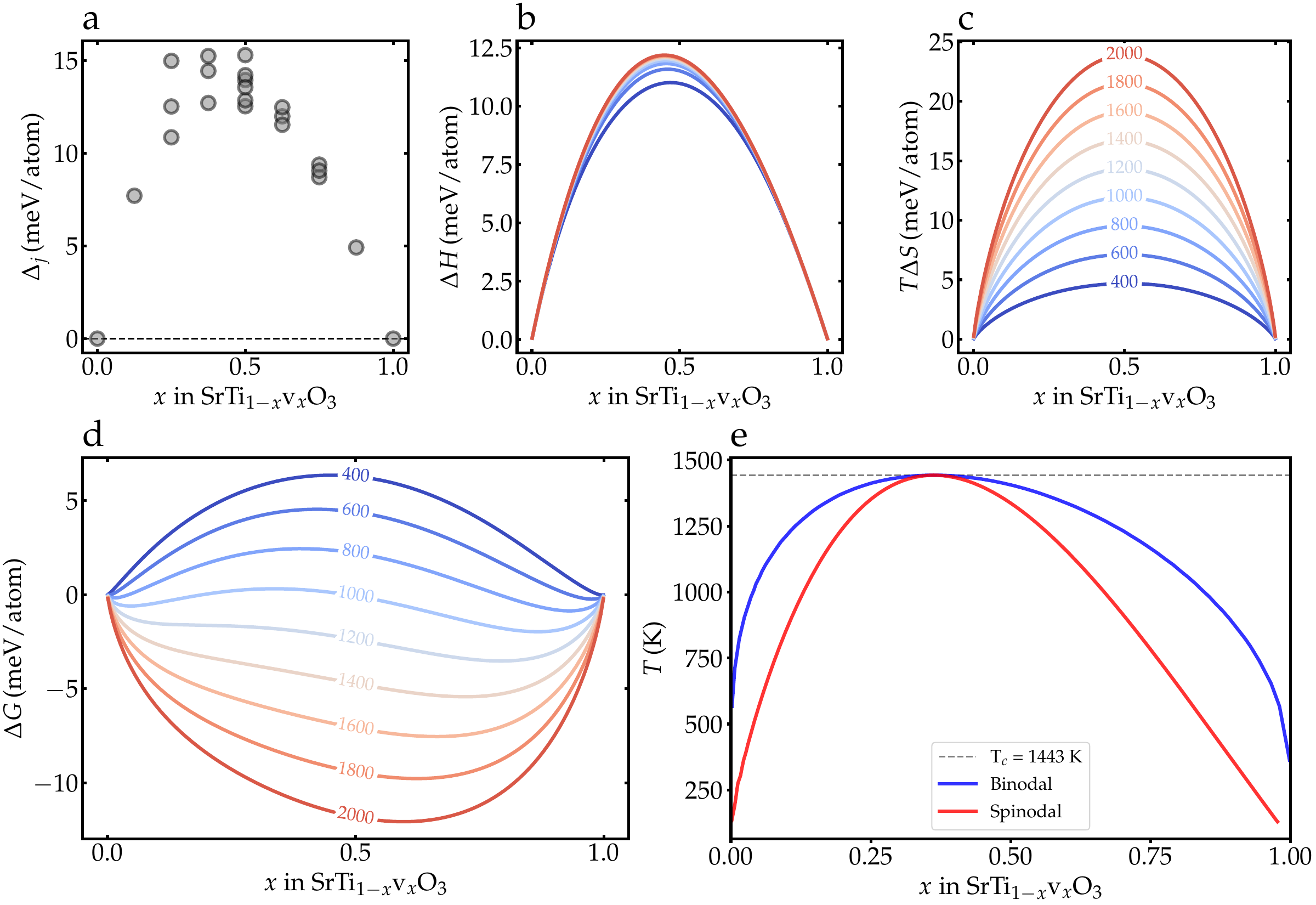}
    \caption{Thermodynamic properties of SrTi$_{1-x}$V$_x$O$_3$ as a function of composition. (a) Excess enthalpies $\Delta_j$. (b) Mixing enthalpy $\Delta H$, (c)  configurational entropy $T\Delta S$ and (d) Gibbs mixing free energy $\Delta G$ at different Temperatures from 400 K to 2000 K. (e) The miscibility gap with binodal and spinodal lines.}
    \label{fig:thermodynamics}
\end{figure*}

\textit{Thermodynamic modeling.}
Although our calculated lattice parameters obey Vegard's law, suggesting the formation of an ideal random solid solution, the actual microscopic distribution of atoms is governed by the balance between the internal energy of mixing and the entropy, which together determine the minimum free energy configuration. To evaluate the mixing thermodynamics of SrTiO$_3$ and SrVO$_3$, we present the GQCA-derived thermodynamic quantities in Fig.~\ref{fig:thermodynamics}. Fig.~\ref{fig:thermodynamics}(a) shows the excess energy $\Delta_j$, which measures the stability of each off-stoichiometric ordered configuration relative to the end-members. All the 22 configurations considered in the $2\times2\times2$ supercell exhibit positive excess energies, indicating that none of them are energetically favored with respect to phase separation into the end-member compounds. Consequently, the mixing enthalpy $\Delta H$ remains positive over the entire composition and temperature range.

Under these conditions, the free energy is lowered by increasing the number of Ti--Ti and V--V nearest-neighbor bonds, favoring the formation of Ti-rich and V-rich regions. This indicates that phase separation is thermodynamically preferred at low temperatures in this system. 

Nevertheless, above a finite critical temperature $T_{\text{c}}$, the configurational entropy should become sufficiently large to stabilize the pseudo-binary alloy over the full composition range. Below $T_{\text{c}}$, however, the mixing free energy should develop a concave region, leading to a miscibility gap. The equilibrium compositions that delimit this two-phase region are obtained from the common-tangent construction applied to the mixing free-energy curve. These tangent points form the binodal boundaries of the miscibility gap.

To analyze this behavior, we plotted the Gibbs mixing free energy $\Delta G$ of SrTi$_{1-x}$V$_x$O$_3$ as a function of composition at different temperatures, Fig.~\ref{fig:thermodynamics}(d). Below the 1500\,K isothermal line, a common tangent can be constructed between two distinct compositions, indicating that a homogeneous solid solution is no longer the equilibrium state within this composition interval. Instead, the system lowers its free energy by separating into Ti-rich and V-rich phases whose equilibrium compositions are given by the tangent points. 

A second important boundary is the spinodal curve, which is defined by the inflection points of the free-energy curve:
\begin{align}
\left. \dfrac{\partial^2 G(x,T)}{\partial x^2} \right|_{x=x_{1}^{\mathrm{'}}}
=
\left. \dfrac{\partial^2 G(x,T)}{\partial x^2} \right|_{x=x_{2}^{\mathrm{'}}}
= 0 .
\end{align}
The spinodal curve separates metastable and unstable regions inside the miscibility gap. For compositions inside the spinodal region, $x_{1}^{\mathrm{'}} < x < x_{2}^{\mathrm{'}}$, the curvature of the free energy is negative, $\partial^2 G/\partial x^2 < 0$. In this case, even infinitesimal composition fluctuations lower the free energy, and the homogeneous alloy is unstable against spinodal decomposition into Ti-rich and V-rich regions.

By contrast, compositions located between the binodal and spinodal curves correspond to a metastable homogeneous alloy. In this region, phase separation is thermodynamically favored, but it must begin through the nucleation of a Ti-rich or V-rich region within the surrounding matrix. Forming such a region creates an interface, which costs energy. For very small nuclei, this interfacial energy cost is larger than the bulk free-energy gain from phase separation, and the nuclei are therefore unstable. Once a nucleus exceeds a critical size, however, the free-energy gain becomes large enough to overcome the interfacial penalty, allowing the nucleus to grow. Phase separation in this metastable region therefore occurs through a nucleation-and-growth mechanism, governed by cation diffusion and the associated kinetic barriers.

Fig.~\ref{fig:thermodynamics}(e) shows the calculated miscibility gap for SrTi$_{1-x}$V$_x$O$_3$. We predict a critical temperature of 1443\,K. Experimental information on the phase equilibria of this system is rather limited. As far as we could identify, no thermodynamic assessment or phase diagram has been reported for the Sr--Ti--V--O quaternary system. SrTi$_{1-x}$V$_x$O$_3$ solid solutions have been synthesized using several different methods, including powders treated at 1250\,K under flowing H$_2$ followed by hydrostatic pressing~\cite{tsuiki1983}, solid-state synthesis of pellets heated at $T\approx 1800$\,K~\cite{hong2002}, and different thin-film growth techniques~\cite{itaki2005,gu2013,kanda2021}.

However, Patino \emph{et al.}~\cite{patino2018} reported limited solubility between SrVO$_3$ and SrTiO$_3$ under ceramic processing conditions. In particular, they observed that quenched SrTi$_{1-x}$V$_x$O$_3$ samples annealed at 1100\,$^\circ$C fully decomposed into simple mixtures of SrVO$_3$ and SrTiO$_3$, in excellent agreement with our predicted miscibility gap. They also reported an increased SrTiO$_3$--SrVO$_3$ solubility in samples prepared by arc melting. They therefore concluded that the SrTi$_{1-x}$V$_x$O$_3$ solid solution is stable only at high temperature, but can be partially retained as a metastable phase upon rapid cooling.

In addition to our thermodynamic assessment, Liu \emph{et al.}~\cite{liu2018} also predicted the formation of a miscibility gap using cluster expansion and Monte Carlo simulations. In their work, a GGA+$U$ correction was applied to the V-$d$ orbitals. However, their calculations substantially underestimated the miscibility gap, predicting a critical temperature around 500\,K, well below the temperature range in which experimental evidence of phase separation is observed. This comparison highlights the robustness of the GQCA approach for modeling the mixing thermodynamics. To the best of our knowledge, the present work provides the first fully \emph{ab initio} prediction of a miscibility gap in SrTi$_{1-x}$V$_x$O$_3$ that is consistent with the available experimental evidence.

It's important to note that the binodal and spinodal curves describe the thermodynamic driving force for phase separation, but they do not determine how fast this process occurs in a real sample. The kinetics are mainly controlled by the mobility of Ti and V cations, which must diffuse through the lattice to form Ti-rich and V-rich domains. At high temperatures, cation diffusion is accelerated, and the system can more readily approach the equilibrium phase-separated state predicted by the free-energy curves. At lower temperatures, however, cation mobility may be strongly reduced, allowing the homogeneous solid solution to persist even when phase separation is thermodynamically favorable. In this case, the alloy is kinetically trapped in a metastable state because Ti and V atoms cannot redistribute efficiently on the experimental time scale. Thus, the calculated miscibility gap reflects the equilibrium tendency of SrTi$_{1-x}$V$_x$O$_3$ to decompose into Ti-rich and V-rich phases, whereas the experimentally observed microstructure will also depend on diffusion kinetics, synthesis temperature, annealing time, and thermal history.

\begin{figure}[t]
    \centering
    \includegraphics[width=0.98\linewidth]{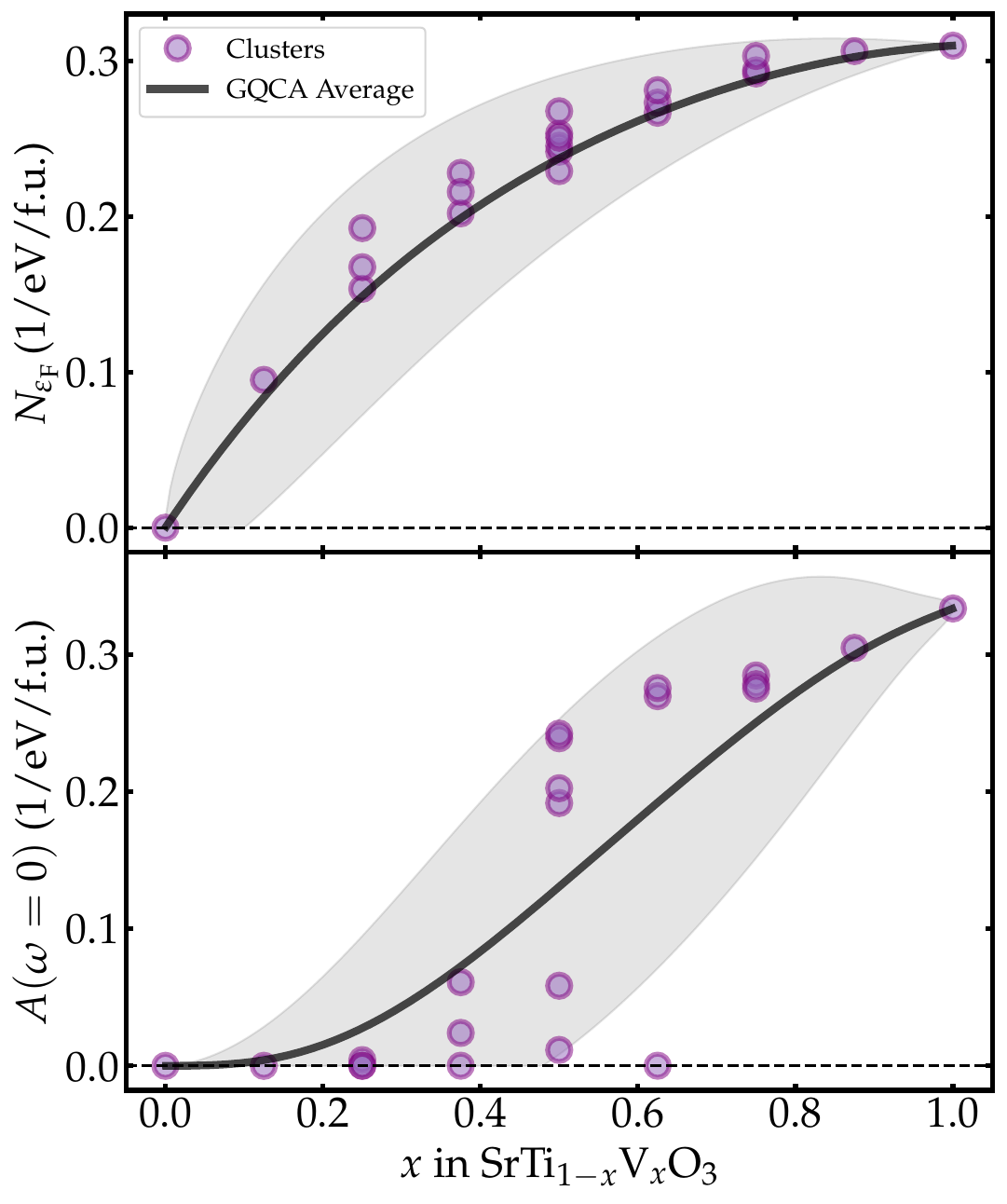}
    \caption{Electronic density of states at the Fermi level taken from the DFT calculation and the spectral function $A(\omega)$ at $\omega=0$ of SrTi$_{1-x}$V$_x$O$_3$ as a function of composition. Both the GQCA average and the individual cluster values are plotted. The gray shaded area shows the GQCA standard deviation.}
    \label{fig:nef_A0}
\end{figure}

\textit{DOS at the Fermi level.}
We now turn to the electronic properties of the SrTi$_{1-x}$V$_x$O$_3$ system. A first simple measure of the electronic properties might be the density of states at the Fermi level obtained from the DFT calculations alone. As can be seen in Fig. \ref{fig:nef_A0}, these show a non-zero value and therefore metallic properties for all compositions $x > 0$. This clearly contradicts experimental evidence of a metal--insulator transition between $x=0.4$ and $x=0.7$~\cite{tsuiki1983,hong2002,gu2013,itaki2005,kanda2021}. This absence of a MIT at the DFT level highlights the need to include correlation effects to capture the Mott physics driving the transition -- consistent with earlier supercell calculations at selected compositions~\cite{James_2024}. We incorporate these effects through DMFT.
        
\begin{figure}[th]
    \centering
    \includegraphics[width=0.9\linewidth]{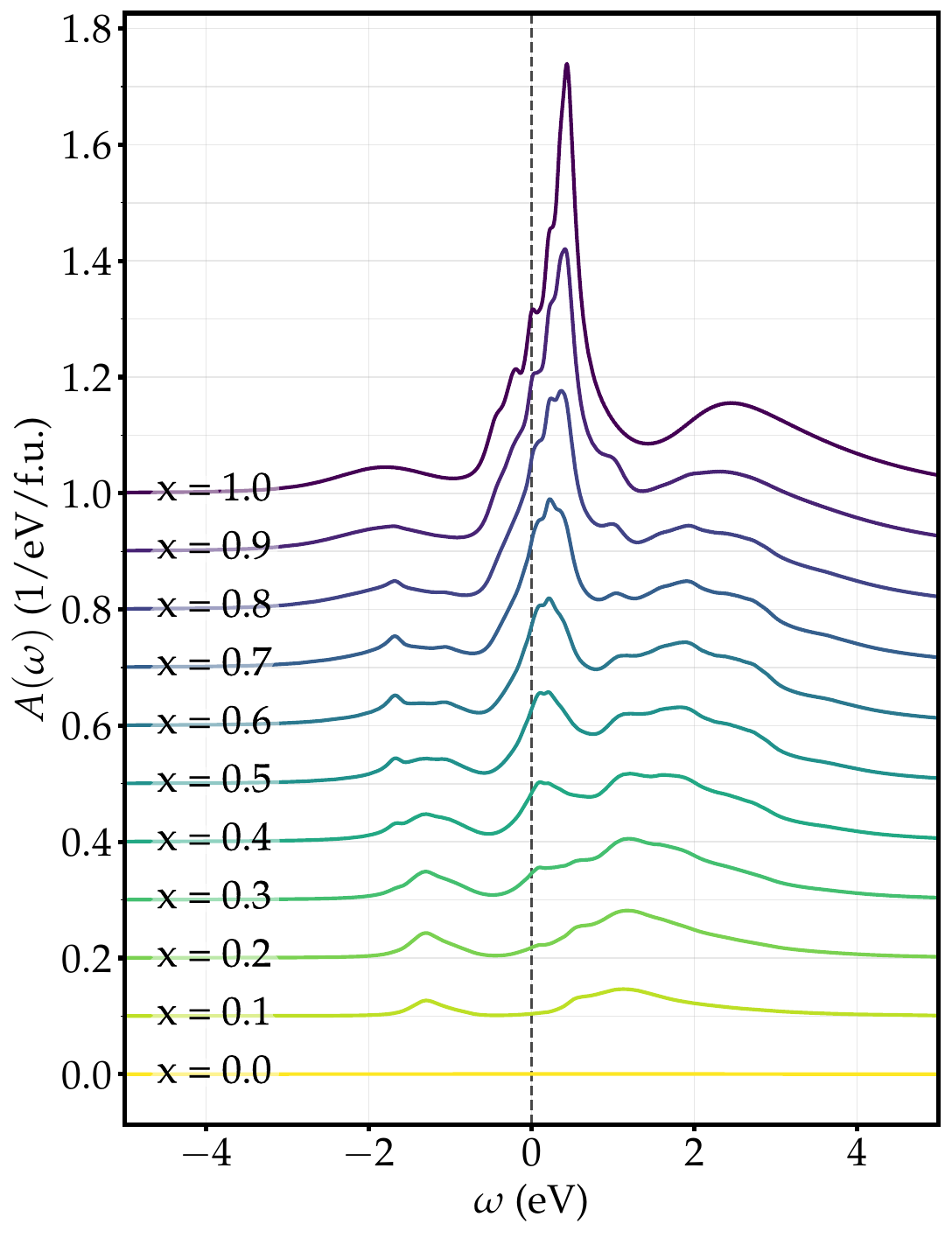}
    \caption{Vanadium-projected GQCA averaged spectral function $A(\omega)$ for different compositions $x$ in SrTi$_{1-x}$V$_x$O$_3$. Each spectral function is shifted vertically by 0.1 eV$^{-1}$ for better visibility. Note that only the vanadium-projected part is shown, that is why for $x=0.0$ the spectral function is identically zero.}
    \label{fig:A_sigma}
\end{figure}

\textit{The spectral function.} To assess whether the inclusion of correlation effects provides an improved description of the MIT, we examine the DMFT analogue of the Fermi-level DOS, namely the spectral function at zero frequency, $A(\omega=0)$. This quantity is obtained directly from the DMFT Green's function $G(\tau)$, as described in Eq.~\ref{eq:A_0}, and is shown in Fig.~\ref{fig:nef_A0}. In contrast to the DFT DOS at the Fermi level, all clusters with $x < 0.3$ exhibit vanishing $A(\omega=0)$, indicating the absence of free charge carriers and hence insulating behavior. Around $x = 0.5$, both metallic and insulating clusters are found. Consequently, the GQCA average displays a sigmoid-like transition from an insulating state at low $x$ to a metallic state as $x$ approaches unity.

Clusters with compositions near $x=0.5$ exhibit pronounced variations in $A(\omega=0)$, ranging from fully insulating behavior to values approaching those of pure SrVO$_3$. This finding highlights the central role of local configurational effects in the MIT: clusters with identical composition can lie on opposite sides of the transition solely as a consequence of the specific arrangement of V and Ti sites. A similar effect was previously reported by James \emph{et al.}~\cite{James_2024}, who concluded that the MIT arises from the combined influence of local configurational effects and Mott physics.

An important ingredient is that the local configuration around the V atoms controls the bandwidth of the relevant correlated states, as shown in Fig.~\ref{fig:dos_vasp_selected} in the Appendix. Since the Coulomb interactions remain essentially unchanged, being purely local and only weakly affected by the surrounding environment, a reduction of the bandwidth drives the system toward stronger electronic correlations and ultimately insulating behavior.

GQCA is well suited to this situation: it averages over all cluster configurations weighted by their occurrence probabilities, thereby retaining the local effects that govern the MIT while simultaneously capturing the thermodynamic solid solution realized at finite temperature on a much larger scale.

Looking at the full spectral function obtained by a maximum entropy analytic continuation of the DMFT self-energy depicted in Fig. \ref{fig:A_sigma}, we can see a clear peak around $\omega=0$ for the metallic compositions. This peak then opens up into a band gap starting at compositions below $x = 0.3$. Thus we have the typical behavior of a Mott transition dependent on the composition $x$, with a metallic peak at $x=1$, a reduced metallic peak and both upper and lower Hubbard bands for $x\approx0.5$ and insulating behavior with a band gap at $x\approx0.1$.

\begin{figure}[t]
    \centering
    \includegraphics[width=\linewidth]{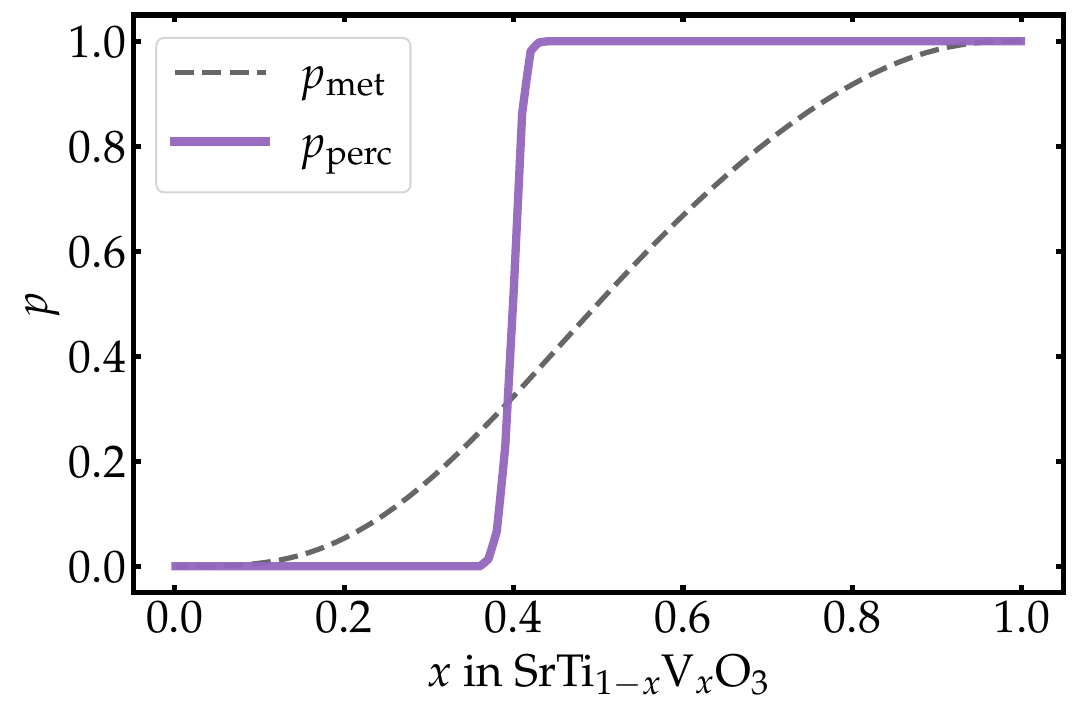}
    \caption{Probability of percolation $p_\mathrm{perc}$ at compositions $x$ in SrTi$_{1-x}$V$_x$O$_3$. Percolation is measured for a 3-dimensional system, from one side of the sample to the other. Additionally the probability that a site is metallic $p_\mathrm{met}$ is plotted over the composition $x$.}
    \label{fig:p_perc_met}
\end{figure}

\textit{A percolation picture.}
So far we have characterized the transition
through the spectral function $A(\omega=0)$, a local quantity that is
naturally suited to the cluster-based thermodynamic treatment of GQCA. The
distinction between a metal and an insulator, however, is ultimately tied to
whether charge can flow across the macroscopic sample, i.e.\ whether a
connected conducting path spans the system. In the following, we adopt this
complementary, transport-oriented point of view.

To this end, we assign a binary character to each cluster: for simplicity, we
call clusters whose spectral function lies below a threshold value
$A_\mathrm{c} = 0.1$\,(eV$\cdot$f.u.)$^{-1}$ insulating, and all others
metallic. Since the GQCA minimization already provides the occurrence
probability $p_j$ of every cluster, the probability that a given site is
metallic follows as the combined weight of the metallic clusters,
\begin{equation}
    p_\mathrm{met}(x,T) = \sum_{j \,\in\, \mathrm{met}} p_j(x,T)
    \label{eq:p_met}
\end{equation}
with the insulating fraction given by $p_\mathrm{ins}(x,T) = 1 - p_\mathrm{met}(x,T)$. We again used the temperature $T=1800$\,K.

We then use $p_\mathrm{met}(x)$ as the site-occupation probability in a
site-percolation simulation on a three-dimensional simple cubic lattice, and
extract the probability $p_\mathrm{perc}(x)$ that the metallic sites form a
cluster that spans the system. The result is shown in
Fig.~\ref{fig:p_perc_met}. Under these admittedly crude assumptions,
$p_\mathrm{perc}(x)$ still exhibits a clear transition: the system is
insulating ($p_\mathrm{perc} \approx 0$) at small $x$ and turns metallic
($p_\mathrm{perc} \approx 1$) at $x = 0.40$. This onset is consistent
with the site-percolation threshold of the simple cubic lattice,
$p_c = 0.3116$~\cite{deng2005}, which the metallic fraction $p_\mathrm{met}(x)$
reaches at a comparable composition. The percolation analysis therefore places
the metal--insulator transition in the same composition window as the
spectral-function analysis, now interpreted as the emergence of a
system-spanning conducting network.

Changing the threshold used to classify metallic clusters to $A_\mathrm{c} = 0.03$~(eV$\cdot$f.u.)$^{-1}$, which identifies two additional clusters as metallic, shifts the estimated metal--insulator transition composition downward by only $0.04$, to $x = 0.36$. This demonstrates that the transition composition is relatively robust with respect to the chosen metal--insulator threshold $A_\mathrm{c}$.

\section{Conclusion}

In conclusion, we have adapted a method from alloy modelling, the GQCA, to the
composition-driven Mott transition in SrTi$_{1-x}$V$_x$O$_3$. Previously, such MIT 
calculations were carried out using individual supercell
configurations~\cite{James_2024}, which has the shortcoming of having to select
individual clusters as representative of a given composition. Here, instead, we
employ a statistical, thermodynamically consistent scheme that overcomes this
limitation.

As a result, we find that the miscibility limit lies at very high
temperatures, which explains the limited experimental access to the
solid-solution phase. Furthermore, we consistently find a critical composition for the MIT
of around $x \approx0.4$, in agreement with the available
experimental data. Importantly, this transition is driven by changes in the
local environment of the vanadium atoms. Starting from the clean SrVO$_3$ end
member, the relevant bandwidth of the localised V $t_{2g}$ states shrinks
significantly as V is replaced by Ti, whereas the local interaction strength is
expected to remain largely unchanged. As a consequence, the local Coulomb
interaction becomes far more effective on the SrTiO$_3$-rich side of the
composition range and drives the system through a metal--insulator transition.
This picture is corroborated by a percolation study of the alloy: we computed
the occupation probability of finding a metallic cluster in the system and used
it as input for a three-dimensional percolation simulation, which again yields a
transition around $x \approx0.4$.

Beyond SrTi$_{1-x}$V$_x$O$_3$, the framework presented here is relevant to the
much broader class of materials in which substitutional disorder governs the
physical properties and can even drive electronic transitions. 
A defining feature of SrTi$_{1-x}$V$_x$O$_3$ is that the substitution acts
directly on the states at the Fermi level: V carries the $t_{2g}$ electron that
forms these states, whereas the $3d^0$ Ti site is electronically inert, so
alloying controls both the number and the spatial connectivity of the
orbital-active sites. This places the system in a broad class of correlated
materials in which substitution reshapes the Fermi-level electronic structure
itself, rather than acting only indirectly through chemical pressure or carrier
doping. In the iron pnictides, isovalent Ru-for-Fe substitution in
BaFe$_2$As$_2$ weakens the electronic correlations and enlarges the
Fermi-surface pockets without adding carriers~\cite{brouet2010}.
More generally, even an individual substitutional defect can locally
redistribute the Fermi-level spectral weight and induce magnetism in a
correlated host~\cite{alloul2009}. A thermodynamically consistent,
configuration-averaged framework such as the GQCA is well suited to
capture these effects and could provide a unifying, fully \emph{ab initio}
perspective on this wider family of disordered correlated systems.


\section*{Author Contributions}
LW performed the \emph{ab initio} calculations. 
LW, PNF, and MA wrote the first draft.
LW, DS, and PNF developed the GQCA code. 
CO carried out the percolation calculation.
PNF and MA supervised this project. 
All authors participated in the discussions and revised the manuscript.

\begin{acknowledgments}
MA thanks insightful discussions with Alyn James and Jude Laverock.
This research has been funded by the Austrian Science Fund (FWF) under projects  
DOI 10.55776/PAT1157425 (LW and CH), 10.55776/ESP8588124 (PNF), and 10.55776/PAT1184524 (MA).

\section*{Data availability}

The ab-initio calculations have been performed using VASP in  version 6.3 as well as TRIQS version 3.3. Codes and scripts, as well as the data shown in this work, are publicly available on a permanent repository~\cite{data}.

\end{acknowledgments}

\appendix
\begin{figure*}[ht]
    \centering
    \includegraphics[width=0.95\textwidth]{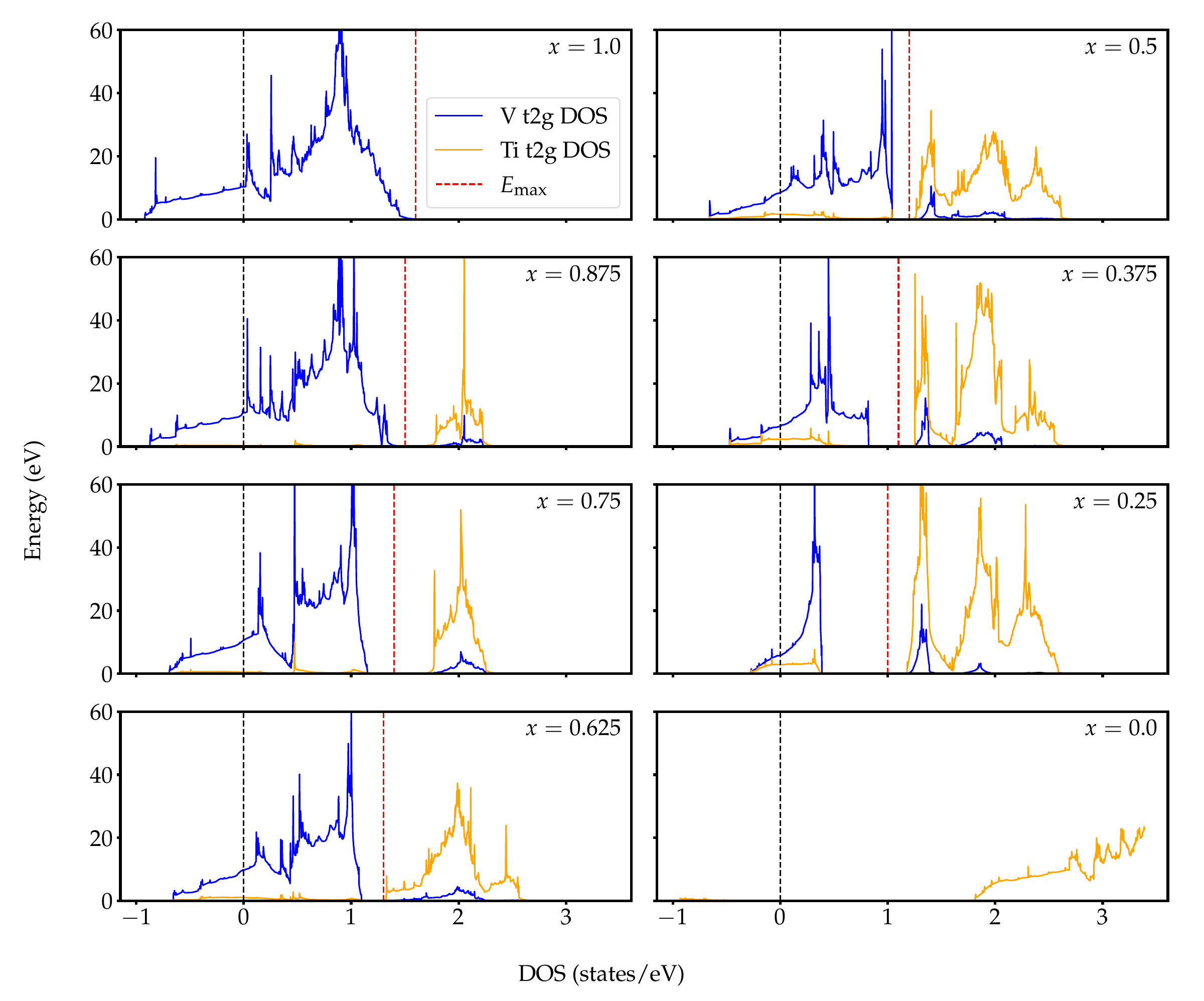}
    \caption{Partial density of states for the $t_{2g}$ orbitals of selected $2\times2\times2$ SrTi$_{1-x}$V$_x$O$_3$ clusters with compositions $x$. Note that here $x$ refers to the composition of specific $2\times2\times2$ clusters, not the GQCA average at $x$. The upper bound of the Wannier projection $E_\mathrm{max}$ is drawn as red dashed vertical line, while the lower bound is always $E_\mathrm{min}$ = -1.44 eV. For pure SrTiO$_3$ ($x=0.0$), there are no states at the Fermi level, and we do not construct a low energy model.}
    \label{fig:dos_vasp_selected}
\end{figure*}

\section{Density of states of selected supercells}
In this appendix, we show density of states for some selected supercells, for different compositions. Starting from the clean SrVO$_3$ example, where the bandwidth is $W_{x=1}=2.52$\,eV, it diminishes continuously to $W_{x=0.25} = 0.67$\,eV. Although not visible directly in this plot, the low-energy density of states around the Fermi level always consists of the three $t_{2g}$ bands from vanadium. At the clean SrTiO$_3$ side, i.e. for $x=0$, the $d$-shell is empty and the system is a band insulator, and we do not construct a low-energy model in that case.

\bibliography{references}   

\begin{thebibliography}{50}%
\makeatletter
\providecommand \@ifxundefined [1]{%
 \@ifx{#1\undefined}
}%
\providecommand \@ifnum [1]{%
 \ifnum #1\expandafter \@firstoftwo
 \else \expandafter \@secondoftwo
 \fi
}%
\providecommand \@ifx [1]{%
 \ifx #1\expandafter \@firstoftwo
 \else \expandafter \@secondoftwo
 \fi
}%
\providecommand \natexlab [1]{#1}%
\providecommand \enquote  [1]{``#1''}%
\providecommand \bibnamefont  [1]{#1}%
\providecommand \bibfnamefont [1]{#1}%
\providecommand \citenamefont [1]{#1}%
\providecommand \href@noop [0]{\@secondoftwo}%
\providecommand \href [0]{\begingroup \@sanitize@url \@href}%
\providecommand \@href[1]{\@@startlink{#1}\@@href}%
\providecommand \@@href[1]{\endgroup#1\@@endlink}%
\providecommand \@sanitize@url [0]{\catcode `\\12\catcode `\$12\catcode
  `\&12\catcode `\#12\catcode `\^12\catcode `\_12\catcode `\%12\relax}%
\providecommand \@@startlink[1]{}%
\providecommand \@@endlink[0]{}%
\providecommand \url  [0]{\begingroup\@sanitize@url \@url }%
\providecommand \@url [1]{\endgroup\@href {#1}{\urlprefix }}%
\providecommand \urlprefix  [0]{URL }%
\providecommand \Eprint [0]{\href }%
\providecommand \doibase [0]{https://doi.org/}%
\providecommand \selectlanguage [0]{\@gobble}%
\providecommand \bibinfo  [0]{\@secondoftwo}%
\providecommand \bibfield  [0]{\@secondoftwo}%
\providecommand \translation [1]{[#1]}%
\providecommand \BibitemOpen [0]{}%
\providecommand \bibitemStop [0]{}%
\providecommand \bibitemNoStop [0]{.\EOS\space}%
\providecommand \EOS [0]{\spacefactor3000\relax}%
\providecommand \BibitemShut  [1]{\csname bibitem#1\endcsname}%
\let\auto@bib@innerbib\@empty
\bibitem [{\citenamefont {Imada}\ \emph {et~al.}(1998)\citenamefont {Imada},
  \citenamefont {Fujimori},\ and\ \citenamefont {Tokura}}]{imada1998}%
  \BibitemOpen
  \bibfield  {author} {\bibinfo {author} {\bibfnamefont {M.}~\bibnamefont
  {Imada}}, \bibinfo {author} {\bibfnamefont {A.}~\bibnamefont {Fujimori}},\
  and\ \bibinfo {author} {\bibfnamefont {Y.}~\bibnamefont {Tokura}},\
  }\bibfield  {title} {\bibinfo {title} {Metal-insulator transitions},\ }\href
  {https://doi.org/10.1103/RevModPhys.70.1039} {\bibfield  {journal} {\bibinfo
  {journal} {Rev. Mod. Phys.}\ }\textbf {\bibinfo {volume} {70}},\ \bibinfo
  {pages} {1039} (\bibinfo {year} {1998})}\BibitemShut {NoStop}%
\bibitem [{\citenamefont {Georges}\ \emph {et~al.}(1996)\citenamefont
  {Georges}, \citenamefont {Kotliar}, \citenamefont {Krauth},\ and\
  \citenamefont {Rozenberg}}]{georges1996}%
  \BibitemOpen
  \bibfield  {author} {\bibinfo {author} {\bibfnamefont {A.}~\bibnamefont
  {Georges}}, \bibinfo {author} {\bibfnamefont {G.}~\bibnamefont {Kotliar}},
  \bibinfo {author} {\bibfnamefont {W.}~\bibnamefont {Krauth}},\ and\ \bibinfo
  {author} {\bibfnamefont {M.~J.}\ \bibnamefont {Rozenberg}},\ }\bibfield
  {title} {\bibinfo {title} {Dynamical mean-field theory of strongly correlated
  fermion systems and the limit of infinite dimensions},\ }\href
  {https://doi.org/10.1103/RevModPhys.68.13} {\bibfield  {journal} {\bibinfo
  {journal} {Rev. Mod. Phys.}\ }\textbf {\bibinfo {volume} {68}},\ \bibinfo
  {pages} {13} (\bibinfo {year} {1996})}\BibitemShut {NoStop}%
\bibitem [{\citenamefont {Tokura}\ and\ \citenamefont
  {Nagaosa}(2000)}]{tokura2000}%
  \BibitemOpen
  \bibfield  {author} {\bibinfo {author} {\bibfnamefont {Y.}~\bibnamefont
  {Tokura}}\ and\ \bibinfo {author} {\bibfnamefont {N.}~\bibnamefont
  {Nagaosa}},\ }\bibfield  {title} {\bibinfo {title} {Orbital physics in
  transition-metal oxides},\ }\href
  {https://doi.org/10.1126/science.288.5465.462} {\bibfield  {journal}
  {\bibinfo  {journal} {Science}\ }\textbf {\bibinfo {volume} {288}},\ \bibinfo
  {pages} {462} (\bibinfo {year} {2000})}\BibitemShut {NoStop}%
\bibitem [{\citenamefont {Ha}\ and\ \citenamefont {Ramanathan}(2011)}]{ha2011}%
  \BibitemOpen
  \bibfield  {author} {\bibinfo {author} {\bibfnamefont {S.~D.}\ \bibnamefont
  {Ha}}\ and\ \bibinfo {author} {\bibfnamefont {S.}~\bibnamefont
  {Ramanathan}},\ }\bibfield  {title} {\bibinfo {title} {Adaptive oxide
  electronics: A review},\ }\href {https://doi.org/10.1063/1.3640806}
  {\bibfield  {journal} {\bibinfo  {journal} {Journal of Applied Physics}\
  }\textbf {\bibinfo {volume} {110}},\ \bibinfo {pages} {071101} (\bibinfo
  {year} {2011})}\BibitemShut {NoStop}%
\bibitem [{\citenamefont {Coll}\ \emph {et~al.}(2019)\citenamefont {Coll},
  \citenamefont {Fontcuberta}, \citenamefont {Althammer}, \citenamefont {Bibes}
  \emph {et~al.}}]{coll2019}%
  \BibitemOpen
  \bibfield  {author} {\bibinfo {author} {\bibfnamefont {M.}~\bibnamefont
  {Coll}}, \bibinfo {author} {\bibfnamefont {J.}~\bibnamefont {Fontcuberta}},
  \bibinfo {author} {\bibfnamefont {M.}~\bibnamefont {Althammer}}, \bibinfo
  {author} {\bibfnamefont {M.}~\bibnamefont {Bibes}}, \emph {et~al.},\
  }\bibfield  {title} {\bibinfo {title} {Towards oxide electronics: a
  roadmap},\ }\href {https://doi.org/10.1016/j.apsusc.2019.03.312} {\bibfield
  {journal} {\bibinfo  {journal} {Applied Surface Science}\ }\textbf {\bibinfo
  {volume} {482}},\ \bibinfo {pages} {1} (\bibinfo {year} {2019})}\BibitemShut
  {NoStop}%
\bibitem [{\citenamefont {Fujimori}\ \emph {et~al.}(1992)\citenamefont
  {Fujimori}, \citenamefont {Hase}, \citenamefont {Namatame}, \citenamefont
  {Fujishima}, \citenamefont {Tokura}, \citenamefont {Eisaki}, \citenamefont
  {Uchida}, \citenamefont {Takegahara},\ and\ \citenamefont
  {de~Groot}}]{fujimori1992}%
  \BibitemOpen
  \bibfield  {author} {\bibinfo {author} {\bibfnamefont {A.}~\bibnamefont
  {Fujimori}}, \bibinfo {author} {\bibfnamefont {I.}~\bibnamefont {Hase}},
  \bibinfo {author} {\bibfnamefont {H.}~\bibnamefont {Namatame}}, \bibinfo
  {author} {\bibfnamefont {Y.}~\bibnamefont {Fujishima}}, \bibinfo {author}
  {\bibfnamefont {Y.}~\bibnamefont {Tokura}}, \bibinfo {author} {\bibfnamefont
  {H.}~\bibnamefont {Eisaki}}, \bibinfo {author} {\bibfnamefont
  {S.}~\bibnamefont {Uchida}}, \bibinfo {author} {\bibfnamefont
  {K.}~\bibnamefont {Takegahara}},\ and\ \bibinfo {author} {\bibfnamefont
  {F.~M.~F.}\ \bibnamefont {de~Groot}},\ }\bibfield  {title} {\bibinfo {title}
  {{Evolution of the spectral function in Mott-Hubbard systems with $d^1$
  configuration}},\ }\href {https://doi.org/10.1103/PhysRevLett.69.1796}
  {\bibfield  {journal} {\bibinfo  {journal} {Phys. Rev. Lett.}\ }\textbf
  {\bibinfo {volume} {69}},\ \bibinfo {pages} {1796} (\bibinfo {year}
  {1992})}\BibitemShut {NoStop}%
\bibitem [{\citenamefont {Inoue}\ \emph {et~al.}(1995)\citenamefont {Inoue},
  \citenamefont {Hase}, \citenamefont {Aiura}, \citenamefont {Fujimori},
  \citenamefont {Haruyama}, \citenamefont {Maruyama},\ and\ \citenamefont
  {Nishihara}}]{inoue1995}%
  \BibitemOpen
  \bibfield  {author} {\bibinfo {author} {\bibfnamefont {I.~H.}\ \bibnamefont
  {Inoue}}, \bibinfo {author} {\bibfnamefont {I.}~\bibnamefont {Hase}},
  \bibinfo {author} {\bibfnamefont {Y.}~\bibnamefont {Aiura}}, \bibinfo
  {author} {\bibfnamefont {A.}~\bibnamefont {Fujimori}}, \bibinfo {author}
  {\bibfnamefont {Y.}~\bibnamefont {Haruyama}}, \bibinfo {author}
  {\bibfnamefont {T.}~\bibnamefont {Maruyama}},\ and\ \bibinfo {author}
  {\bibfnamefont {Y.}~\bibnamefont {Nishihara}},\ }\bibfield  {title} {\bibinfo
  {title} {{Systematic Development of the Spectral Function in the $3d^1$
  Mott-Hubbard System Ca$_{1-x}$Sr$_x$VO$_3$}},\ }\href
  {https://doi.org/10.1103/PhysRevLett.74.2539} {\bibfield  {journal} {\bibinfo
   {journal} {Phys. Rev. Lett.}\ }\textbf {\bibinfo {volume} {74}},\ \bibinfo
  {pages} {2539} (\bibinfo {year} {1995})}\BibitemShut {NoStop}%
\bibitem [{\citenamefont {Tsuiki}\ \emph {et~al.}(1983)\citenamefont {Tsuiki},
  \citenamefont {Kitazawa},\ and\ \citenamefont {Fueki}}]{tsuiki1983}%
  \BibitemOpen
  \bibfield  {author} {\bibinfo {author} {\bibfnamefont {H.}~\bibnamefont
  {Tsuiki}}, \bibinfo {author} {\bibfnamefont {K.}~\bibnamefont {Kitazawa}},\
  and\ \bibinfo {author} {\bibfnamefont {K.}~\bibnamefont {Fueki}},\ }\bibfield
   {title} {\bibinfo {title} {The donor level of {V4+} and the {Metal-Nonmetal}
  transition in {SrTi$_{1-x}$V$_x$O$_3$}},\ }\href
  {https://doi.org/10.1143/JJAP.22.590} {\bibfield  {journal} {\bibinfo
  {journal} {Japanese Journal of Applied Physics}\ }\textbf {\bibinfo {volume}
  {22}},\ \bibinfo {pages} {590} (\bibinfo {year} {1983})}\BibitemShut
  {NoStop}%
\bibitem [{\citenamefont {Hong}\ \emph {et~al.}(2002)\citenamefont {Hong},
  \citenamefont {Kim}, \citenamefont {Heo},\ and\ \citenamefont
  {Kwon}}]{hong2002}%
  \BibitemOpen
  \bibfield  {author} {\bibinfo {author} {\bibfnamefont {K.}~\bibnamefont
  {Hong}}, \bibinfo {author} {\bibfnamefont {S.-H.}\ \bibnamefont {Kim}},
  \bibinfo {author} {\bibfnamefont {Y.-J.}\ \bibnamefont {Heo}},\ and\ \bibinfo
  {author} {\bibfnamefont {Y.-U.}\ \bibnamefont {Kwon}},\ }\bibfield  {title}
  {\bibinfo {title} {Metal–insulator transitions of {SrTi$_{1-x}$V$_x$O$_3$}
  solid solution system},\ }\href
  {https://doi.org/https://doi.org/10.1016/S0038-1098(02)00297-1} {\bibfield
  {journal} {\bibinfo  {journal} {Solid State Communications}\ }\textbf
  {\bibinfo {volume} {123}},\ \bibinfo {pages} {305} (\bibinfo {year}
  {2002})}\BibitemShut {NoStop}%
\bibitem [{\citenamefont {Gu}\ \emph {et~al.}(2013)\citenamefont {Gu},
  \citenamefont {Wolf},\ and\ \citenamefont {Lu}}]{gu2013}%
  \BibitemOpen
  \bibfield  {author} {\bibinfo {author} {\bibfnamefont {M.}~\bibnamefont
  {Gu}}, \bibinfo {author} {\bibfnamefont {S.~A.}\ \bibnamefont {Wolf}},\ and\
  \bibinfo {author} {\bibfnamefont {J.}~\bibnamefont {Lu}},\ }\bibfield
  {title} {\bibinfo {title} {Metal-insulator transition in
  {SrTi$_{1-x}$V$_x$O$_3$} thin films},\ }\href
  {https://doi.org/10.1063/1.4836576} {\bibfield  {journal} {\bibinfo
  {journal} {Applied Physics Letters}\ }\textbf {\bibinfo {volume} {103}},\
  \bibinfo {pages} {223110} (\bibinfo {year} {2013})}\BibitemShut {NoStop}%
\bibitem [{\citenamefont {Kanda}\ \emph {et~al.}(2021)\citenamefont {Kanda},
  \citenamefont {Shiga}, \citenamefont {Yukawa}, \citenamefont {Hasegawa},
  \citenamefont {Nguyen}, \citenamefont {Cheng}, \citenamefont {Tokunaga},
  \citenamefont {Kitamura}, \citenamefont {Horiba}, \citenamefont
  {Yoshimatsu},\ and\ \citenamefont {Kumigashira}}]{kanda2021}%
  \BibitemOpen
  \bibfield  {author} {\bibinfo {author} {\bibfnamefont {T.}~\bibnamefont
  {Kanda}}, \bibinfo {author} {\bibfnamefont {D.}~\bibnamefont {Shiga}},
  \bibinfo {author} {\bibfnamefont {R.}~\bibnamefont {Yukawa}}, \bibinfo
  {author} {\bibfnamefont {N.}~\bibnamefont {Hasegawa}}, \bibinfo {author}
  {\bibfnamefont {D.~K.}\ \bibnamefont {Nguyen}}, \bibinfo {author}
  {\bibfnamefont {X.}~\bibnamefont {Cheng}}, \bibinfo {author} {\bibfnamefont
  {R.}~\bibnamefont {Tokunaga}}, \bibinfo {author} {\bibfnamefont
  {M.}~\bibnamefont {Kitamura}}, \bibinfo {author} {\bibfnamefont
  {K.}~\bibnamefont {Horiba}}, \bibinfo {author} {\bibfnamefont
  {K.}~\bibnamefont {Yoshimatsu}},\ and\ \bibinfo {author} {\bibfnamefont
  {H.}~\bibnamefont {Kumigashira}},\ }\bibfield  {title} {\bibinfo {title}
  {Electronic structure of {SrTi$_{1-x}$V$_{x}$O$_3$} films studied by in situ
  photoemission spectroscopy: Screening for a transparent electrode material},\
  }\href {https://doi.org/10.1103/PhysRevB.104.115121} {\bibfield  {journal}
  {\bibinfo  {journal} {Phys. Rev. B}\ }\textbf {\bibinfo {volume} {104}},\
  \bibinfo {pages} {115121} (\bibinfo {year} {2021})}\BibitemShut {NoStop}%
\bibitem [{\citenamefont {Itaka}\ \emph {et~al.}(2005)\citenamefont {Itaka},
  \citenamefont {Wakisaka}, \citenamefont {Mihara}, \citenamefont {Yaginuma},
  \citenamefont {Matsumoto},\ and\ \citenamefont {Koinuma}}]{itaki2005}%
  \BibitemOpen
  \bibfield  {author} {\bibinfo {author} {\bibfnamefont {K.}~\bibnamefont
  {Itaka}}, \bibinfo {author} {\bibfnamefont {T.}~\bibnamefont {Wakisaka}},
  \bibinfo {author} {\bibfnamefont {T.}~\bibnamefont {Mihara}}, \bibinfo
  {author} {\bibfnamefont {S.}~\bibnamefont {Yaginuma}}, \bibinfo {author}
  {\bibfnamefont {Y.}~\bibnamefont {Matsumoto}},\ and\ \bibinfo {author}
  {\bibfnamefont {H.}~\bibnamefont {Koinuma}},\ }\bibfield  {title} {\bibinfo
  {title} {Sharp metal-insulator transition in
  {Sr(Ti$_{1-x}$V$_x$)O$_3$-$\delta$} thin films on {SrTiO$_3$} substrates},\
  }\href {https://doi.org/https://doi.org/10.1016/j.tsf.2004.11.230} {\bibfield
   {journal} {\bibinfo  {journal} {Thin Solid Films}\ }\textbf {\bibinfo
  {volume} {486}},\ \bibinfo {pages} {222} (\bibinfo {year}
  {2005})}\BibitemShut {NoStop}%
\bibitem [{\citenamefont {James}\ \emph {et~al.}(2024)\citenamefont {James},
  \citenamefont {Aichhorn},\ and\ \citenamefont {Laverock}}]{James_2024}%
  \BibitemOpen
  \bibfield  {author} {\bibinfo {author} {\bibfnamefont {A.~D.~N.}\
  \bibnamefont {James}}, \bibinfo {author} {\bibfnamefont {M.}~\bibnamefont
  {Aichhorn}},\ and\ \bibinfo {author} {\bibfnamefont {J.}~\bibnamefont
  {Laverock}},\ }\bibfield  {title} {\bibinfo {title} {Composition-driven mott
  transition within {SrTi$_{1-x}$V$_x$O$_3$}},\ }\href
  {https://doi.org/10.1088/2516-1075/ad29ab} {\bibfield  {journal} {\bibinfo
  {journal} {Electronic Structure}\ }\textbf {\bibinfo {volume} {6}},\ \bibinfo
  {pages} {015010} (\bibinfo {year} {2024})}\BibitemShut {NoStop}%
\bibitem [{\citenamefont {Laverock}\ \emph {et~al.}(2025)\citenamefont
  {Laverock}, \citenamefont {James}, \citenamefont {Gu}, \citenamefont {Lu},\
  and\ \citenamefont {Smith}}]{laverock2025}%
  \BibitemOpen
  \bibfield  {author} {\bibinfo {author} {\bibfnamefont {J.}~\bibnamefont
  {Laverock}}, \bibinfo {author} {\bibfnamefont {A.~D.~N.}\ \bibnamefont
  {James}}, \bibinfo {author} {\bibfnamefont {M.}~\bibnamefont {Gu}}, \bibinfo
  {author} {\bibfnamefont {J.~W.}\ \bibnamefont {Lu}},\ and\ \bibinfo {author}
  {\bibfnamefont {K.~E.}\ \bibnamefont {Smith}},\ }\bibfield  {title} {\bibinfo
  {title} {{Role of electron--electron correlations and disorder in the
  metal-insulator transition of SrTi$_{1-x}$V$_x$O$_3$ thin films}},\ }\href
  {https://doi.org/10.1088/2516-1075/ae2b40} {\bibfield  {journal} {\bibinfo
  {journal} {Electronic Structure}\ }\textbf {\bibinfo {volume} {7}},\ \bibinfo
  {pages} {045003} (\bibinfo {year} {2025})}\BibitemShut {NoStop}%
\bibitem [{\citenamefont {Liu}\ \emph {et~al.}(2018)\citenamefont {Liu},
  \citenamefont {Podraza}, \citenamefont {Khare},\ and\ \citenamefont
  {Sarin}}]{liu2018}%
  \BibitemOpen
  \bibfield  {author} {\bibinfo {author} {\bibfnamefont {Z.}~\bibnamefont
  {Liu}}, \bibinfo {author} {\bibfnamefont {N.}~\bibnamefont {Podraza}},
  \bibinfo {author} {\bibfnamefont {S.}~\bibnamefont {Khare}},\ and\ \bibinfo
  {author} {\bibfnamefont {P.}~\bibnamefont {Sarin}},\ }\bibfield  {title}
  {\bibinfo {title} {Transparency enhancement for {SrVO$_3$} by {SrTiO$_3$}
  mixing: A first-principles study},\ }\href
  {https://doi.org/https://doi.org/10.1016/j.commatsci.2017.12.020} {\bibfield
  {journal} {\bibinfo  {journal} {Computational Materials Science}\ }\textbf
  {\bibinfo {volume} {144}},\ \bibinfo {pages} {139} (\bibinfo {year}
  {2018})}\BibitemShut {NoStop}%
\bibitem [{\citenamefont {Anisimov}\ \emph {et~al.}(1991)\citenamefont
  {Anisimov}, \citenamefont {Zaanen},\ and\ \citenamefont
  {Andersen}}]{anisimov1991}%
  \BibitemOpen
  \bibfield  {author} {\bibinfo {author} {\bibfnamefont {V.~I.}\ \bibnamefont
  {Anisimov}}, \bibinfo {author} {\bibfnamefont {J.}~\bibnamefont {Zaanen}},\
  and\ \bibinfo {author} {\bibfnamefont {O.~K.}\ \bibnamefont {Andersen}},\
  }\bibfield  {title} {\bibinfo {title} {{Band theory and Mott insulators:
  Hubbard U instead of Stoner I}},\ }\href
  {https://doi.org/10.1103/PhysRevB.44.943} {\bibfield  {journal} {\bibinfo
  {journal} {Phys. Rev. B}\ }\textbf {\bibinfo {volume} {44}},\ \bibinfo
  {pages} {943} (\bibinfo {year} {1991})}\BibitemShut {NoStop}%
\bibitem [{\citenamefont {Dudarev}\ \emph {et~al.}(1998)\citenamefont
  {Dudarev}, \citenamefont {Botton}, \citenamefont {Savrasov}, \citenamefont
  {Humphreys},\ and\ \citenamefont {Sutton}}]{dudarev1998}%
  \BibitemOpen
  \bibfield  {author} {\bibinfo {author} {\bibfnamefont {S.~L.}\ \bibnamefont
  {Dudarev}}, \bibinfo {author} {\bibfnamefont {G.~A.}\ \bibnamefont {Botton}},
  \bibinfo {author} {\bibfnamefont {S.~Y.}\ \bibnamefont {Savrasov}}, \bibinfo
  {author} {\bibfnamefont {C.~J.}\ \bibnamefont {Humphreys}},\ and\ \bibinfo
  {author} {\bibfnamefont {A.~P.}\ \bibnamefont {Sutton}},\ }\bibfield  {title}
  {\bibinfo {title} {{Electron-energy-loss spectra and the structural stability
  of nickel oxide: An LSDA+U study}},\ }\href
  {https://doi.org/10.1103/PhysRevB.57.1505} {\bibfield  {journal} {\bibinfo
  {journal} {Phys. Rev. B}\ }\textbf {\bibinfo {volume} {57}},\ \bibinfo
  {pages} {1505} (\bibinfo {year} {1998})}\BibitemShut {NoStop}%
\bibitem [{\citenamefont {Sher}\ \emph {et~al.}(1987)\citenamefont {Sher},
  \citenamefont {van Schilfgaarde}, \citenamefont {Chen},\ and\ \citenamefont
  {Chen}}]{sher1987GQCA}%
  \BibitemOpen
  \bibfield  {author} {\bibinfo {author} {\bibfnamefont {A.}~\bibnamefont
  {Sher}}, \bibinfo {author} {\bibfnamefont {M.}~\bibnamefont {van
  Schilfgaarde}}, \bibinfo {author} {\bibfnamefont {A.-B.}\ \bibnamefont
  {Chen}},\ and\ \bibinfo {author} {\bibfnamefont {W.}~\bibnamefont {Chen}},\
  }\bibfield  {title} {\bibinfo {title} {Quasichemical approximation in binary
  alloys},\ }\href {https://doi.org/10.1103/PhysRevB.36.4279} {\bibfield
  {journal} {\bibinfo  {journal} {Phys. Rev. B}\ }\textbf {\bibinfo {volume}
  {36}},\ \bibinfo {pages} {4279} (\bibinfo {year} {1987})}\BibitemShut
  {NoStop}%
\bibitem [{\citenamefont {Nunes~Ferreira}\ \emph {et~al.}(2024)\citenamefont
  {Nunes~Ferreira}, \citenamefont {Lucrezi}, \citenamefont {Guilhon},
  \citenamefont {Marques}, \citenamefont {Teles}, \citenamefont {Heil},\ and\
  \citenamefont {Eleno}}]{ferreira2024EGQCA}%
  \BibitemOpen
  \bibfield  {author} {\bibinfo {author} {\bibfnamefont {P.}~\bibnamefont
  {Nunes~Ferreira}}, \bibinfo {author} {\bibfnamefont {R.}~\bibnamefont
  {Lucrezi}}, \bibinfo {author} {\bibfnamefont {I.}~\bibnamefont {Guilhon}},
  \bibinfo {author} {\bibfnamefont {M.}~\bibnamefont {Marques}}, \bibinfo
  {author} {\bibfnamefont {L.}~\bibnamefont {Teles}}, \bibinfo {author}
  {\bibfnamefont {C.}~\bibnamefont {Heil}},\ and\ \bibinfo {author}
  {\bibfnamefont {L.}~\bibnamefont {Eleno}},\ }\bibfield  {title} {\bibinfo
  {title} {Ab initio modeling of superconducting alloys},\ }\href
  {https://doi.org/10.48550/arXiv.2406.15174} {\bibfield  {journal} {\bibinfo
  {journal} {Materials Today Physics}\ }\textbf {\bibinfo {volume} {48}},\
  \bibinfo {pages} {101547} (\bibinfo {year} {2024})}\BibitemShut {NoStop}%
\bibitem [{\citenamefont {Spath}\ \emph {et~al.}(2026)\citenamefont {Spath},
  \citenamefont {Wibmer}, \citenamefont {Meyer}, \citenamefont {Di~Cataldo},
  \citenamefont {Madeddu}, \citenamefont {Aichhorn}, \citenamefont {Eleno},
  \citenamefont {Heil},\ and\ \citenamefont {Nunes~Ferreira}}]{EGQCA_code}%
  \BibitemOpen
  \bibfield  {author} {\bibinfo {author} {\bibfnamefont {D.}~\bibnamefont
  {Spath}}, \bibinfo {author} {\bibfnamefont {L.}~\bibnamefont {Wibmer}},
  \bibinfo {author} {\bibfnamefont {R.}~\bibnamefont {Meyer}}, \bibinfo
  {author} {\bibfnamefont {S.}~\bibnamefont {Di~Cataldo}}, \bibinfo {author}
  {\bibfnamefont {E.}~\bibnamefont {Madeddu}}, \bibinfo {author} {\bibfnamefont
  {M.}~\bibnamefont {Aichhorn}}, \bibinfo {author} {\bibfnamefont
  {L.}~\bibnamefont {Eleno}}, \bibinfo {author} {\bibfnamefont
  {C.}~\bibnamefont {Heil}},\ and\ \bibinfo {author} {\bibfnamefont
  {P.}~\bibnamefont {Nunes~Ferreira}},\ }\bibfield  {title} {\bibinfo {title}
  {{EGQCA Code: streamlining first-principles simulations of real-world
  alloys}},\ }\href@noop {} {\bibfield  {journal} {\bibinfo  {journal}
  {Unpublished}\ } (\bibinfo {year} {2026})}\BibitemShut {NoStop}%
\bibitem [{\citenamefont {Amano~Patino}\ \emph {et~al.}(2018)\citenamefont
  {Amano~Patino}, \citenamefont {Zeng}, \citenamefont {Blundell}, \citenamefont
  {McGrady},\ and\ \citenamefont {Hayward}}]{patino2018}%
  \BibitemOpen
  \bibfield  {author} {\bibinfo {author} {\bibfnamefont {M.}~\bibnamefont
  {Amano~Patino}}, \bibinfo {author} {\bibfnamefont {D.}~\bibnamefont {Zeng}},
  \bibinfo {author} {\bibfnamefont {S.~J.}\ \bibnamefont {Blundell}}, \bibinfo
  {author} {\bibfnamefont {J.~E.}\ \bibnamefont {McGrady}},\ and\ \bibinfo
  {author} {\bibfnamefont {M.~A.}\ \bibnamefont {Hayward}},\ }\bibfield
  {title} {\bibinfo {title} {Extreme sensitivity of a topochemical reaction to
  cation substitution: {SrVO$_2$H} versus
  {SrV$_{1–x}$Ti$_x$O$_{1.5}$H$_{1.5}$}},\ }\href
  {https://doi.org/10.1021/acs.inorgchem.8b00026} {\bibfield  {journal}
  {\bibinfo  {journal} {Inorganic Chemistry}\ }\textbf {\bibinfo {volume}
  {57}},\ \bibinfo {pages} {2890} (\bibinfo {year} {2018})}\BibitemShut
  {NoStop}%
\bibitem [{\citenamefont {Stauffer}\ and\ \citenamefont
  {Aharony}(1994)}]{stauffer1994}%
  \BibitemOpen
  \bibfield  {author} {\bibinfo {author} {\bibfnamefont {D.}~\bibnamefont
  {Stauffer}}\ and\ \bibinfo {author} {\bibfnamefont {A.}~\bibnamefont
  {Aharony}},\ }\href@noop {} {\emph {\bibinfo {title} {Introduction to
  Percolation Theory}}},\ \bibinfo {edition} {revised 2nd}\ ed.\ (\bibinfo
  {publisher} {Taylor \& Francis},\ \bibinfo {address} {London},\ \bibinfo
  {year} {1994})\BibitemShut {NoStop}%
\bibitem [{\citenamefont {Deng}\ and\ \citenamefont
  {Bl{\"o}te}(2005)}]{deng2005}%
  \BibitemOpen
  \bibfield  {author} {\bibinfo {author} {\bibfnamefont {Y.}~\bibnamefont
  {Deng}}\ and\ \bibinfo {author} {\bibfnamefont {H.~W.~J.}\ \bibnamefont
  {Bl{\"o}te}},\ }\bibfield  {title} {\bibinfo {title} {Monte carlo study of
  the site-percolation model in two and three dimensions},\ }\href
  {https://doi.org/10.1103/PhysRevE.72.016126} {\bibfield  {journal} {\bibinfo
  {journal} {Phys. Rev. E}\ }\textbf {\bibinfo {volume} {72}},\ \bibinfo
  {pages} {016126} (\bibinfo {year} {2005})}\BibitemShut {NoStop}%
\bibitem [{\citenamefont {Teles}\ \emph {et~al.}(2000)\citenamefont {Teles},
  \citenamefont {Furthm\"uller}, \citenamefont {Scolfaro}, \citenamefont
  {Leite},\ and\ \citenamefont {Bechstedt}}]{teles2000GQCA}%
  \BibitemOpen
  \bibfield  {author} {\bibinfo {author} {\bibfnamefont {L.~K.}\ \bibnamefont
  {Teles}}, \bibinfo {author} {\bibfnamefont {J.}~\bibnamefont
  {Furthm\"uller}}, \bibinfo {author} {\bibfnamefont {L.~M.~R.}\ \bibnamefont
  {Scolfaro}}, \bibinfo {author} {\bibfnamefont {J.~R.}\ \bibnamefont
  {Leite}},\ and\ \bibinfo {author} {\bibfnamefont {F.}~\bibnamefont
  {Bechstedt}},\ }\bibfield  {title} {\bibinfo {title} {First-principles
  calculations of the thermodynamic and structural properties of strained
  {${\mathrm{In}}_{x}{\mathrm{Ga}}_{1\ensuremath{-}x}\mathrm{N}$} and
  {${\mathrm{Al}}_{x}{\mathrm{Ga}}_{1\ensuremath{-}x}\mathrm{N}$} alloys},\
  }\href {https://doi.org/10.1103/PhysRevB.62.2475} {\bibfield  {journal}
  {\bibinfo  {journal} {Phys. Rev. B}\ }\textbf {\bibinfo {volume} {62}},\
  \bibinfo {pages} {2475} (\bibinfo {year} {2000})}\BibitemShut {NoStop}%
\bibitem [{\citenamefont {Guilhon}\ \emph {et~al.}(2015)\citenamefont
  {Guilhon}, \citenamefont {Teles}, \citenamefont {Marques}, \citenamefont
  {Pelá},\ and\ \citenamefont {Bechstedt}}]{guilhon2015GQCA}%
  \BibitemOpen
  \bibfield  {author} {\bibinfo {author} {\bibfnamefont {I.}~\bibnamefont
  {Guilhon}}, \bibinfo {author} {\bibfnamefont {L.}~\bibnamefont {Teles}},
  \bibinfo {author} {\bibfnamefont {M.}~\bibnamefont {Marques}}, \bibinfo
  {author} {\bibfnamefont {R.}~\bibnamefont {Pelá}},\ and\ \bibinfo {author}
  {\bibfnamefont {F.}~\bibnamefont {Bechstedt}},\ }\bibfield  {title} {\bibinfo
  {title} {{Influence of structure and thermodynamic stability on electronic
  properties of two-dimensional SiC, SiGe, and GeC alloys}},\ }\bibfield
  {journal} {\bibinfo  {journal} {Physical Review B}\ }\textbf {\bibinfo
  {volume} {92}},\ \href {https://doi.org/10.1103/PhysRevB.92.075435}
  {10.1103/PhysRevB.92.075435} (\bibinfo {year} {2015})\BibitemShut {NoStop}%
\bibitem [{\citenamefont {Guilhon}\ \emph {et~al.}(2017)\citenamefont
  {Guilhon}, \citenamefont {Bechstedt}, \citenamefont {Botti}, \citenamefont
  {Marques},\ and\ \citenamefont {Teles}}]{guilhon2017GQCA}%
  \BibitemOpen
  \bibfield  {author} {\bibinfo {author} {\bibfnamefont {I.}~\bibnamefont
  {Guilhon}}, \bibinfo {author} {\bibfnamefont {F.}~\bibnamefont {Bechstedt}},
  \bibinfo {author} {\bibfnamefont {S.}~\bibnamefont {Botti}}, \bibinfo
  {author} {\bibfnamefont {M.}~\bibnamefont {Marques}},\ and\ \bibinfo {author}
  {\bibfnamefont {L.~K.}\ \bibnamefont {Teles}},\ }\bibfield  {title} {\bibinfo
  {title} {Thermodynamic, electronic, and optical properties of graphene oxide:
  A statistical ab initio approach},\ }\href
  {https://doi.org/10.1103/PhysRevB.95.245427} {\bibfield  {journal} {\bibinfo
  {journal} {Phys. Rev. B}\ }\textbf {\bibinfo {volume} {95}},\ \bibinfo
  {pages} {245427} (\bibinfo {year} {2017})}\BibitemShut {NoStop}%
\bibitem [{\citenamefont {Okhotnikov}\ \emph {et~al.}(2016)\citenamefont
  {Okhotnikov}, \citenamefont {Charpentier},\ and\ \citenamefont
  {Cadars}}]{supercell}%
  \BibitemOpen
  \bibfield  {author} {\bibinfo {author} {\bibfnamefont {K.}~\bibnamefont
  {Okhotnikov}}, \bibinfo {author} {\bibfnamefont {T.}~\bibnamefont
  {Charpentier}},\ and\ \bibinfo {author} {\bibfnamefont {S.}~\bibnamefont
  {Cadars}},\ }\bibfield  {title} {\bibinfo {title} {Supercell program: A
  combinatorial structure-generation approach for the local-level modeling of
  atomic substitutions and partial occupancies in crystals},\ }\bibfield
  {journal} {\bibinfo  {journal} {Journal of Cheminformatics}\ }\textbf
  {\bibinfo {volume} {8}},\ \href {https://doi.org/10.1186/s13321-016-0129-3}
  {10.1186/s13321-016-0129-3} (\bibinfo {year} {2016})\BibitemShut {NoStop}%
\bibitem [{\citenamefont {Kresse}\ and\ \citenamefont
  {Joubert}(1999)}]{vasp_paw}%
  \BibitemOpen
  \bibfield  {author} {\bibinfo {author} {\bibfnamefont {G.}~\bibnamefont
  {Kresse}}\ and\ \bibinfo {author} {\bibfnamefont {D.}~\bibnamefont
  {Joubert}},\ }\bibfield  {title} {\bibinfo {title} {From ultrasoft
  pseudopotentials to the projector augmented-wave method},\ }\href
  {https://doi.org/10.1103/PhysRevB.59.1758} {\bibfield  {journal} {\bibinfo
  {journal} {Phys. Rev. B}\ }\textbf {\bibinfo {volume} {59}},\ \bibinfo
  {pages} {1758} (\bibinfo {year} {1999})}\BibitemShut {NoStop}%
\bibitem [{\citenamefont {Bl\"ochl}(1994)}]{PAW_method}%
  \BibitemOpen
  \bibfield  {author} {\bibinfo {author} {\bibfnamefont {P.~E.}\ \bibnamefont
  {Bl\"ochl}},\ }\bibfield  {title} {\bibinfo {title} {Projector augmented-wave
  method},\ }\href {https://doi.org/10.1103/PhysRevB.50.17953} {\bibfield
  {journal} {\bibinfo  {journal} {Phys. Rev. B}\ }\textbf {\bibinfo {volume}
  {50}},\ \bibinfo {pages} {17953} (\bibinfo {year} {1994})}\BibitemShut
  {NoStop}%
\bibitem [{\citenamefont {Parcollet}\ \emph {et~al.}(2015)\citenamefont
  {Parcollet}, \citenamefont {Ferrero}, \citenamefont {Ayral}, \citenamefont
  {Hafermann}, \citenamefont {Krivenko}, \citenamefont {Messio},\ and\
  \citenamefont {Seth}}]{TRIQS}%
  \BibitemOpen
  \bibfield  {author} {\bibinfo {author} {\bibfnamefont {O.}~\bibnamefont
  {Parcollet}}, \bibinfo {author} {\bibfnamefont {M.}~\bibnamefont {Ferrero}},
  \bibinfo {author} {\bibfnamefont {T.}~\bibnamefont {Ayral}}, \bibinfo
  {author} {\bibfnamefont {H.}~\bibnamefont {Hafermann}}, \bibinfo {author}
  {\bibfnamefont {I.}~\bibnamefont {Krivenko}}, \bibinfo {author}
  {\bibfnamefont {L.}~\bibnamefont {Messio}},\ and\ \bibinfo {author}
  {\bibfnamefont {P.}~\bibnamefont {Seth}},\ }\bibfield  {title} {\bibinfo
  {title} {{TRIQS: A toolbox for research on interacting quantum systems}},\
  }\href {https://doi.org/https://doi.org/10.1016/j.cpc.2015.04.023} {\bibfield
   {journal} {\bibinfo  {journal} {Computer Physics Communications}\ }\textbf
  {\bibinfo {volume} {196}},\ \bibinfo {pages} {398} (\bibinfo {year}
  {2015})}\BibitemShut {NoStop}%
\bibitem [{\citenamefont {Aichhorn}\ \emph {et~al.}(2016)\citenamefont
  {Aichhorn}, \citenamefont {Pourovskii}, \citenamefont {Seth}, \citenamefont
  {Vildosola}, \citenamefont {Zingl}, \citenamefont {Peil}, \citenamefont
  {Deng}, \citenamefont {Mravlje}, \citenamefont {Kraberger}, \citenamefont
  {Martins}, \citenamefont {Ferrero},\ and\ \citenamefont
  {Parcollet}}]{DFTTools}%
  \BibitemOpen
  \bibfield  {author} {\bibinfo {author} {\bibfnamefont {M.}~\bibnamefont
  {Aichhorn}}, \bibinfo {author} {\bibfnamefont {L.}~\bibnamefont
  {Pourovskii}}, \bibinfo {author} {\bibfnamefont {P.}~\bibnamefont {Seth}},
  \bibinfo {author} {\bibfnamefont {V.}~\bibnamefont {Vildosola}}, \bibinfo
  {author} {\bibfnamefont {M.}~\bibnamefont {Zingl}}, \bibinfo {author}
  {\bibfnamefont {O.~E.}\ \bibnamefont {Peil}}, \bibinfo {author}
  {\bibfnamefont {X.}~\bibnamefont {Deng}}, \bibinfo {author} {\bibfnamefont
  {J.}~\bibnamefont {Mravlje}}, \bibinfo {author} {\bibfnamefont {G.~J.}\
  \bibnamefont {Kraberger}}, \bibinfo {author} {\bibfnamefont {C.}~\bibnamefont
  {Martins}}, \bibinfo {author} {\bibfnamefont {M.}~\bibnamefont {Ferrero}},\
  and\ \bibinfo {author} {\bibfnamefont {O.}~\bibnamefont {Parcollet}},\
  }\bibfield  {title} {\bibinfo {title} {{TRIQS/DFTTools: A TRIQS application
  for ab initio calculations of correlated materials}},\ }\href
  {https://doi.org/https://doi.org/10.1016/j.cpc.2016.03.014} {\bibfield
  {journal} {\bibinfo  {journal} {Computer Physics Communications}\ }\textbf
  {\bibinfo {volume} {204}},\ \bibinfo {pages} {200} (\bibinfo {year}
  {2016})}\BibitemShut {NoStop}%
\bibitem [{\citenamefont {Lechermann}\ \emph {et~al.}(2006)\citenamefont
  {Lechermann}, \citenamefont {Georges}, \citenamefont {Poteryaev},
  \citenamefont {Biermann}, \citenamefont {Posternak}, \citenamefont
  {Yamasaki},\ and\ \citenamefont {Andersen}}]{Lechermann2006}%
  \BibitemOpen
  \bibfield  {author} {\bibinfo {author} {\bibfnamefont {F.}~\bibnamefont
  {Lechermann}}, \bibinfo {author} {\bibfnamefont {A.}~\bibnamefont {Georges}},
  \bibinfo {author} {\bibfnamefont {A.}~\bibnamefont {Poteryaev}}, \bibinfo
  {author} {\bibfnamefont {S.}~\bibnamefont {Biermann}}, \bibinfo {author}
  {\bibfnamefont {M.}~\bibnamefont {Posternak}}, \bibinfo {author}
  {\bibfnamefont {A.}~\bibnamefont {Yamasaki}},\ and\ \bibinfo {author}
  {\bibfnamefont {O.~K.}\ \bibnamefont {Andersen}},\ }\bibfield  {title}
  {\bibinfo {title} {{Dynamical mean-field theory using Wannier functions: A
  flexible route to electronic structure calculations of strongly correlated
  materials}},\ }\href {https://doi.org/10.1103/PhysRevB.74.125120} {\bibfield
  {journal} {\bibinfo  {journal} {Phys. Rev. B}\ }\textbf {\bibinfo {volume}
  {74}},\ \bibinfo {pages} {125120} (\bibinfo {year} {2006})}\BibitemShut
  {NoStop}%
\bibitem [{\citenamefont {Aichhorn}\ \emph {et~al.}(2009)\citenamefont
  {Aichhorn}, \citenamefont {Pourovskii}, \citenamefont {Vildosola},
  \citenamefont {Ferrero}, \citenamefont {Parcollet}, \citenamefont {Miyake},
  \citenamefont {Georges},\ and\ \citenamefont {Biermann}}]{aichhorn1010}%
  \BibitemOpen
  \bibfield  {author} {\bibinfo {author} {\bibfnamefont {M.}~\bibnamefont
  {Aichhorn}}, \bibinfo {author} {\bibfnamefont {L.}~\bibnamefont
  {Pourovskii}}, \bibinfo {author} {\bibfnamefont {V.}~\bibnamefont
  {Vildosola}}, \bibinfo {author} {\bibfnamefont {M.}~\bibnamefont {Ferrero}},
  \bibinfo {author} {\bibfnamefont {O.}~\bibnamefont {Parcollet}}, \bibinfo
  {author} {\bibfnamefont {T.}~\bibnamefont {Miyake}}, \bibinfo {author}
  {\bibfnamefont {A.}~\bibnamefont {Georges}},\ and\ \bibinfo {author}
  {\bibfnamefont {S.}~\bibnamefont {Biermann}},\ }\bibfield  {title} {\bibinfo
  {title} {{Dynamical mean-field theory within an augmented plane-wave
  framework: Assessing electronic correlations in the iron pnictide LaFeAsO}},\
  }\href {https://doi.org/10.1103/PhysRevB.80.085101} {\bibfield  {journal}
  {\bibinfo  {journal} {Phys. Rev. B}\ }\textbf {\bibinfo {volume} {80}},\
  \bibinfo {pages} {085101} (\bibinfo {year} {2009})}\BibitemShut {NoStop}%
\bibitem [{\citenamefont {Vaugier}\ \emph {et~al.}(2012)\citenamefont
  {Vaugier}, \citenamefont {Jiang},\ and\ \citenamefont
  {Biermann}}]{vaugier2012}%
  \BibitemOpen
  \bibfield  {author} {\bibinfo {author} {\bibfnamefont {L.}~\bibnamefont
  {Vaugier}}, \bibinfo {author} {\bibfnamefont {H.}~\bibnamefont {Jiang}},\
  and\ \bibinfo {author} {\bibfnamefont {S.}~\bibnamefont {Biermann}},\
  }\bibfield  {title} {\bibinfo {title} {{Hubbard $U$ and Hund exchange $J$ in
  transition metal oxides: Screening versus localization trends from
  constrained random phase approximation}},\ }\href
  {https://doi.org/10.1103/PhysRevB.86.165105} {\bibfield  {journal} {\bibinfo
  {journal} {Phys. Rev. B}\ }\textbf {\bibinfo {volume} {86}},\ \bibinfo
  {pages} {165105} (\bibinfo {year} {2012})}\BibitemShut {NoStop}%
\bibitem [{\citenamefont {Held}(2007)}]{DC}%
  \BibitemOpen
  \bibfield  {author} {\bibinfo {author} {\bibfnamefont {K.}~\bibnamefont
  {Held}},\ }\bibfield  {title} {\bibinfo {title} {Electronic structure
  calculations using dynamical mean field theory},\ }\href
  {https://doi.org/10.1080/00018730701619647} {\bibfield  {journal} {\bibinfo
  {journal} {Advances in Physics}\ }\textbf {\bibinfo {volume} {56}},\ \bibinfo
  {pages} {829} (\bibinfo {year} {2007})},\ \Eprint
  {https://arxiv.org/abs/https://doi.org/10.1080/00018730701619647}
  {https://doi.org/10.1080/00018730701619647} \BibitemShut {NoStop}%
\bibitem [{\citenamefont {Kavokine}\ \emph {et~al.}(2025)\citenamefont
  {Kavokine}, \citenamefont {Lu}, \citenamefont {Ayral}, \citenamefont
  {Ferrero}, \citenamefont {Wentzell},\ and\ \citenamefont
  {Parcollet}}]{CTSEG}%
  \BibitemOpen
  \bibfield  {author} {\bibinfo {author} {\bibfnamefont {N.}~\bibnamefont
  {Kavokine}}, \bibinfo {author} {\bibfnamefont {H.}~\bibnamefont {Lu}},
  \bibinfo {author} {\bibfnamefont {T.}~\bibnamefont {Ayral}}, \bibinfo
  {author} {\bibfnamefont {M.}~\bibnamefont {Ferrero}}, \bibinfo {author}
  {\bibfnamefont {N.}~\bibnamefont {Wentzell}},\ and\ \bibinfo {author}
  {\bibfnamefont {O.}~\bibnamefont {Parcollet}},\ }\bibfield  {title} {\bibinfo
  {title} {{CTSEG: A segment picture quantum impurity solver based on TRIQS}},\
  }\href {https://doi.org/10.21105/joss.07425} {\bibfield  {journal} {\bibinfo
  {journal} {Journal of Open Source Software}\ }\textbf {\bibinfo {volume}
  {10}},\ \bibinfo {pages} {7425} (\bibinfo {year} {2025})}\BibitemShut
  {NoStop}%
\bibitem [{\citenamefont {Kraberger}\ \emph {et~al.}(2017)\citenamefont
  {Kraberger}, \citenamefont {Triebl}, \citenamefont {Zingl},\ and\
  \citenamefont {Aichhorn}}]{maxent}%
  \BibitemOpen
  \bibfield  {author} {\bibinfo {author} {\bibfnamefont {G.~J.}\ \bibnamefont
  {Kraberger}}, \bibinfo {author} {\bibfnamefont {R.}~\bibnamefont {Triebl}},
  \bibinfo {author} {\bibfnamefont {M.}~\bibnamefont {Zingl}},\ and\ \bibinfo
  {author} {\bibfnamefont {M.}~\bibnamefont {Aichhorn}},\ }\bibfield  {title}
  {\bibinfo {title} {{Maximum entropy formalism for the analytic continuation
  of matrix-valued Green's functions}},\ }\href
  {https://doi.org/10.1103/PhysRevB.96.155128} {\bibfield  {journal} {\bibinfo
  {journal} {Phys. Rev. B}\ }\textbf {\bibinfo {volume} {96}},\ \bibinfo
  {pages} {155128} (\bibinfo {year} {2017})}\BibitemShut {NoStop}%
\bibitem [{\citenamefont {Yamanaka}\ \emph {et~al.}(2002)\citenamefont
  {Yamanaka}, \citenamefont {Hirai},\ and\ \citenamefont
  {Komatsu}}]{yamanaka2002}%
  \BibitemOpen
  \bibfield  {author} {\bibinfo {author} {\bibfnamefont {T.}~\bibnamefont
  {Yamanaka}}, \bibinfo {author} {\bibfnamefont {N.}~\bibnamefont {Hirai}},\
  and\ \bibinfo {author} {\bibfnamefont {Y.}~\bibnamefont {Komatsu}},\
  }\bibfield  {title} {\bibinfo {title} {Structure change of
  {Ca$_{1-x}$Sr$_x$TiO$_3$} perovskite with composition and pressure},\ }\href
  {https://doi.org/10.2138/am-2002-8-917} {\bibfield  {journal} {\bibinfo
  {journal} {American Mineralogist}\ }\textbf {\bibinfo {volume} {87}},\
  \bibinfo {pages} {1183} (\bibinfo {year} {2002})}\BibitemShut {NoStop}%
\bibitem [{\citenamefont {Lan}\ \emph {et~al.}(2003)\citenamefont {Lan},
  \citenamefont {Chen},\ and\ \citenamefont {He}}]{lan2003}%
  \BibitemOpen
  \bibfield  {author} {\bibinfo {author} {\bibfnamefont {Y.}~\bibnamefont
  {Lan}}, \bibinfo {author} {\bibfnamefont {X.}~\bibnamefont {Chen}},\ and\
  \bibinfo {author} {\bibfnamefont {M.}~\bibnamefont {He}},\ }\bibfield
  {title} {\bibinfo {title} {Structure, magnetic susceptibility and resistivity
  properties of {SrVO$_3$}},\ }\href
  {https://doi.org/https://doi.org/10.1016/S0925-8388(02)01349-X} {\bibfield
  {journal} {\bibinfo  {journal} {Journal of Alloys and Compounds}\ }\textbf
  {\bibinfo {volume} {354}},\ \bibinfo {pages} {95} (\bibinfo {year}
  {2003})}\BibitemShut {NoStop}%
\bibitem [{\citenamefont {Denton}\ and\ \citenamefont
  {Ashcroft}(1991)}]{denton1991}%
  \BibitemOpen
  \bibfield  {author} {\bibinfo {author} {\bibfnamefont {A.~R.}\ \bibnamefont
  {Denton}}\ and\ \bibinfo {author} {\bibfnamefont {N.~W.}\ \bibnamefont
  {Ashcroft}},\ }\bibfield  {title} {\bibinfo {title} {Vegard's law},\ }\href
  {https://doi.org/10.1103/PhysRevA.43.3161} {\bibfield  {journal} {\bibinfo
  {journal} {Phys. Rev. A}\ }\textbf {\bibinfo {volume} {43}},\ \bibinfo
  {pages} {3161} (\bibinfo {year} {1991})}\BibitemShut {NoStop}%
\bibitem [{\citenamefont {Macías}\ \emph {et~al.}(2019)\citenamefont
  {Macías}, \citenamefont {Yaremchenko}, \citenamefont
  {Rodríguez-Castellón}, \citenamefont {Starykevich},\ and\ \citenamefont
  {Frade}}]{macias2019}%
  \BibitemOpen
  \bibfield  {author} {\bibinfo {author} {\bibfnamefont {J.}~\bibnamefont
  {Macías}}, \bibinfo {author} {\bibfnamefont {A.~A.}\ \bibnamefont
  {Yaremchenko}}, \bibinfo {author} {\bibfnamefont {E.}~\bibnamefont
  {Rodríguez-Castellón}}, \bibinfo {author} {\bibfnamefont {M.}~\bibnamefont
  {Starykevich}},\ and\ \bibinfo {author} {\bibfnamefont {J.~R.}\ \bibnamefont
  {Frade}},\ }\bibfield  {title} {\bibinfo {title} {Compromising between phase
  stability and electrical performance: {SrVO$_3$–SrTiO$_3$} solid solutions
  as solid oxide fuel cell anode components},\ }\href
  {https://doi.org/https://doi.org/10.1002/cssc.201801727} {\bibfield
  {journal} {\bibinfo  {journal} {ChemSusChem}\ }\textbf {\bibinfo {volume}
  {12}},\ \bibinfo {pages} {240} (\bibinfo {year} {2019})},\ \Eprint
  {https://arxiv.org/abs/https://chemistry-europe.onlinelibrary.wiley.com/doi/pdf/10.1002/cssc.201801727}
  {https://chemistry-europe.onlinelibrary.wiley.com/doi/pdf/10.1002/cssc.201801727}
  \BibitemShut {NoStop}%
\bibitem [{\citenamefont {Perdew}\ \emph {et~al.}(2008)\citenamefont {Perdew},
  \citenamefont {Ruzsinszky}, \citenamefont {Csonka}, \citenamefont {Vydrov},
  \citenamefont {Scuseria}, \citenamefont {Constantin}, \citenamefont {Zhou},\
  and\ \citenamefont {Burke}}]{perdew2008}%
  \BibitemOpen
  \bibfield  {author} {\bibinfo {author} {\bibfnamefont {J.~P.}\ \bibnamefont
  {Perdew}}, \bibinfo {author} {\bibfnamefont {A.}~\bibnamefont {Ruzsinszky}},
  \bibinfo {author} {\bibfnamefont {G.~I.}\ \bibnamefont {Csonka}}, \bibinfo
  {author} {\bibfnamefont {O.~A.}\ \bibnamefont {Vydrov}}, \bibinfo {author}
  {\bibfnamefont {G.~E.}\ \bibnamefont {Scuseria}}, \bibinfo {author}
  {\bibfnamefont {L.~A.}\ \bibnamefont {Constantin}}, \bibinfo {author}
  {\bibfnamefont {X.}~\bibnamefont {Zhou}},\ and\ \bibinfo {author}
  {\bibfnamefont {K.}~\bibnamefont {Burke}},\ }\bibfield  {title} {\bibinfo
  {title} {Restoring the density-gradient expansion for exchange in solids and
  surfaces},\ }\href {https://doi.org/10.1103/PhysRevLett.100.136406}
  {\bibfield  {journal} {\bibinfo  {journal} {Phys. Rev. Lett.}\ }\textbf
  {\bibinfo {volume} {100}},\ \bibinfo {pages} {136406} (\bibinfo {year}
  {2008})}\BibitemShut {NoStop}%
\bibitem [{\citenamefont {Evarestov}\ \emph {et~al.}(2011)\citenamefont
  {Evarestov}, \citenamefont {Blokhin}, \citenamefont {Gryaznov}, \citenamefont
  {Kotomin},\ and\ \citenamefont {Maier}}]{evarestov2011}%
  \BibitemOpen
  \bibfield  {author} {\bibinfo {author} {\bibfnamefont {R.~A.}\ \bibnamefont
  {Evarestov}}, \bibinfo {author} {\bibfnamefont {E.}~\bibnamefont {Blokhin}},
  \bibinfo {author} {\bibfnamefont {D.}~\bibnamefont {Gryaznov}}, \bibinfo
  {author} {\bibfnamefont {E.~A.}\ \bibnamefont {Kotomin}},\ and\ \bibinfo
  {author} {\bibfnamefont {J.}~\bibnamefont {Maier}},\ }\bibfield  {title}
  {\bibinfo {title} {{Phonon calculations in cubic and tetragonal phases of
  SrTiO$_{3}$: A comparative LCAO and plane-wave study}},\ }\href
  {https://doi.org/10.1103/PhysRevB.83.134108} {\bibfield  {journal} {\bibinfo
  {journal} {Phys. Rev. B}\ }\textbf {\bibinfo {volume} {83}},\ \bibinfo
  {pages} {134108} (\bibinfo {year} {2011})}\BibitemShut {NoStop}%
\bibitem [{\citenamefont {Garc\'{\i}a-Fern\'andez}\ \emph
  {et~al.}(2012)\citenamefont {Garc\'{\i}a-Fern\'andez}, \citenamefont {Ghosh},
  \citenamefont {English},\ and\ \citenamefont {Aramburu}}]{garcia2012}%
  \BibitemOpen
  \bibfield  {author} {\bibinfo {author} {\bibfnamefont {P.}~\bibnamefont
  {Garc\'{\i}a-Fern\'andez}}, \bibinfo {author} {\bibfnamefont
  {S.}~\bibnamefont {Ghosh}}, \bibinfo {author} {\bibfnamefont {N.~J.}\
  \bibnamefont {English}},\ and\ \bibinfo {author} {\bibfnamefont {J.~A.}\
  \bibnamefont {Aramburu}},\ }\bibfield  {title} {\bibinfo {title} {Benchmark
  study for the application of density functional theory to the prediction of
  octahedral tilting in perovskites},\ }\href
  {https://doi.org/10.1103/PhysRevB.86.144107} {\bibfield  {journal} {\bibinfo
  {journal} {Phys. Rev. B}\ }\textbf {\bibinfo {volume} {86}},\ \bibinfo
  {pages} {144107} (\bibinfo {year} {2012})}\BibitemShut {NoStop}%
\bibitem [{\citenamefont {Sun}\ \emph {et~al.}(2015)\citenamefont {Sun},
  \citenamefont {Ruzsinszky},\ and\ \citenamefont {Perdew}}]{sun2015}%
  \BibitemOpen
  \bibfield  {author} {\bibinfo {author} {\bibfnamefont {J.}~\bibnamefont
  {Sun}}, \bibinfo {author} {\bibfnamefont {A.}~\bibnamefont {Ruzsinszky}},\
  and\ \bibinfo {author} {\bibfnamefont {J.~P.}\ \bibnamefont {Perdew}},\
  }\bibfield  {title} {\bibinfo {title} {Strongly constrained and appropriately
  normed semilocal density functional},\ }\href
  {https://doi.org/10.1103/PhysRevLett.115.036402} {\bibfield  {journal}
  {\bibinfo  {journal} {Phys. Rev. Lett.}\ }\textbf {\bibinfo {volume} {115}},\
  \bibinfo {pages} {036402} (\bibinfo {year} {2015})}\BibitemShut {NoStop}%
\bibitem [{\citenamefont {Furness}\ \emph {et~al.}(2020)\citenamefont
  {Furness}, \citenamefont {Kaplan}, \citenamefont {Ning}, \citenamefont
  {Perdew},\ and\ \citenamefont {Sun}}]{furness2020}%
  \BibitemOpen
  \bibfield  {author} {\bibinfo {author} {\bibfnamefont {J.~W.}\ \bibnamefont
  {Furness}}, \bibinfo {author} {\bibfnamefont {A.~D.}\ \bibnamefont {Kaplan}},
  \bibinfo {author} {\bibfnamefont {J.}~\bibnamefont {Ning}}, \bibinfo {author}
  {\bibfnamefont {J.~P.}\ \bibnamefont {Perdew}},\ and\ \bibinfo {author}
  {\bibfnamefont {J.}~\bibnamefont {Sun}},\ }\bibfield  {title} {\bibinfo
  {title} {{Accurate and Numerically Efficient r2SCAN Meta-Generalized Gradient
  Approximation}},\ }\href {https://doi.org/10.1021/acs.jpclett.0c02405}
  {\bibfield  {journal} {\bibinfo  {journal} {The Journal of Physical Chemistry
  Letters}\ }\textbf {\bibinfo {volume} {11}},\ \bibinfo {pages} {8208}
  (\bibinfo {year} {2020})},\ \bibinfo {note} {pMID: 32876454},\ \Eprint
  {https://arxiv.org/abs/https://doi.org/10.1021/acs.jpclett.0c02405}
  {https://doi.org/10.1021/acs.jpclett.0c02405} \BibitemShut {NoStop}%
\bibitem [{\citenamefont {Krukau}\ \emph {et~al.}(2006)\citenamefont {Krukau},
  \citenamefont {Vydrov}, \citenamefont {Izmaylov},\ and\ \citenamefont
  {Scuseria}}]{krukau2006}%
  \BibitemOpen
  \bibfield  {author} {\bibinfo {author} {\bibfnamefont {A.~V.}\ \bibnamefont
  {Krukau}}, \bibinfo {author} {\bibfnamefont {O.~A.}\ \bibnamefont {Vydrov}},
  \bibinfo {author} {\bibfnamefont {A.~F.}\ \bibnamefont {Izmaylov}},\ and\
  \bibinfo {author} {\bibfnamefont {G.~E.}\ \bibnamefont {Scuseria}},\
  }\bibfield  {title} {\bibinfo {title} {Influence of the exchange screening
  parameter on the performance of screened hybrid functionals},\ }\href
  {https://doi.org/10.1063/1.2404663} {\bibfield  {journal} {\bibinfo
  {journal} {The Journal of Chemical Physics}\ }\textbf {\bibinfo {volume}
  {125}},\ \bibinfo {pages} {224106} (\bibinfo {year} {2006})}\BibitemShut
  {NoStop}%
\bibitem [{\citenamefont {Brouet}\ \emph {et~al.}(2010)\citenamefont {Brouet},
  \citenamefont {Rullier-Albenque}, \citenamefont {Marsi}, \citenamefont
  {Mansart}, \citenamefont {Aichhorn}, \citenamefont {Biermann}, \citenamefont
  {Faure}, \citenamefont {Perfetti}, \citenamefont {Taleb-Ibrahimi},
  \citenamefont {Le~F{\`e}vre}, \citenamefont {Bertran}, \citenamefont
  {Forget},\ and\ \citenamefont {Colson}}]{brouet2010}%
  \BibitemOpen
  \bibfield  {author} {\bibinfo {author} {\bibfnamefont {V.}~\bibnamefont
  {Brouet}}, \bibinfo {author} {\bibfnamefont {F.}~\bibnamefont
  {Rullier-Albenque}}, \bibinfo {author} {\bibfnamefont {M.}~\bibnamefont
  {Marsi}}, \bibinfo {author} {\bibfnamefont {B.}~\bibnamefont {Mansart}},
  \bibinfo {author} {\bibfnamefont {M.}~\bibnamefont {Aichhorn}}, \bibinfo
  {author} {\bibfnamefont {S.}~\bibnamefont {Biermann}}, \bibinfo {author}
  {\bibfnamefont {J.}~\bibnamefont {Faure}}, \bibinfo {author} {\bibfnamefont
  {L.}~\bibnamefont {Perfetti}}, \bibinfo {author} {\bibfnamefont
  {A.}~\bibnamefont {Taleb-Ibrahimi}}, \bibinfo {author} {\bibfnamefont
  {P.}~\bibnamefont {Le~F{\`e}vre}}, \bibinfo {author} {\bibfnamefont
  {F.}~\bibnamefont {Bertran}}, \bibinfo {author} {\bibfnamefont
  {A.}~\bibnamefont {Forget}},\ and\ \bibinfo {author} {\bibfnamefont
  {D.}~\bibnamefont {Colson}},\ }\bibfield  {title} {\bibinfo {title}
  {{Significant Reduction of Electronic Correlations upon Isovalent Ru
  Substitution of BaFe$_2$As$_2$}},\ }\href
  {https://doi.org/10.1103/PhysRevLett.105.087001} {\bibfield  {journal}
  {\bibinfo  {journal} {Phys. Rev. Lett.}\ }\textbf {\bibinfo {volume} {105}},\
  \bibinfo {pages} {087001} (\bibinfo {year} {2010})}\BibitemShut {NoStop}%
\bibitem [{\citenamefont {Alloul}\ \emph {et~al.}(2009)\citenamefont {Alloul},
  \citenamefont {Bobroff}, \citenamefont {Gabay},\ and\ \citenamefont
  {Hirschfeld}}]{alloul2009}%
  \BibitemOpen
  \bibfield  {author} {\bibinfo {author} {\bibfnamefont {H.}~\bibnamefont
  {Alloul}}, \bibinfo {author} {\bibfnamefont {J.}~\bibnamefont {Bobroff}},
  \bibinfo {author} {\bibfnamefont {M.}~\bibnamefont {Gabay}},\ and\ \bibinfo
  {author} {\bibfnamefont {P.~J.}\ \bibnamefont {Hirschfeld}},\ }\bibfield
  {title} {\bibinfo {title} {Defects in correlated metals and
  superconductors},\ }\href {https://doi.org/10.1103/RevModPhys.81.45}
  {\bibfield  {journal} {\bibinfo  {journal} {Rev. Mod. Phys.}\ }\textbf
  {\bibinfo {volume} {81}},\ \bibinfo {pages} {45} (\bibinfo {year}
  {2009})}\BibitemShut {NoStop}%
\bibitem [{The DOI of the data will be inserted upon final
  publication()}]{data}%
  \BibitemOpen
  The DOI of the data will be inserted upon final publication,\ \href@noop {}
  {}\BibitemShut {NoStop}%
\end{thebibliography}%

\end{document}